\documentclass[useAMS,usenatbib]{mn2e}
\usepackage{float}

\usepackage{savesym}
\usepackage{amsmath}
\savesymbol{iint}
\usepackage{txfonts}
\restoresymbol{TXF}{iint}

\usepackage{subfigure}
\usepackage{epsfig}
\usepackage{enumerate}
\usepackage{url}
\def\simless{\mathbin{\lower 3pt\hbox
{$\rlap{\raise 5pt\hbox{$\char'074$}}\mathchar"7218$}}}   
\def\simmore{\mathbin{\lower 3pt\hbox
{$\rlap{\raise 5pt\hbox{$\char'076$}}\mathchar"7218$}}}   
\newcommand{\be}{\begin{equation}}
\newcommand{\ee}{\end{equation}}
\newcommand       \bea          {\begin{eqnarray}}
\newcommand       \eea          {\end{eqnarray}}
\newcommand       \apj          {ApJ}

\newcommand       \aap          {A\&A}

\def\simlt{\mathrel{\hbox{\rlap{\hbox{\lower4pt\hbox{$\sim$}}}\hbox{$<$}}}}
\def\simgt{\mathrel{\hbox{\rlap{\hbox{\lower4pt\hbox{$\sim$}}}\hbox{$>$}}}}
\def\lesssim{\mathrel{\hbox{\rlap{\hbox{\lower4pt\hbox{$\sim$}}}\hbox{$<$}}}}
\def\gtrsim{\mathrel{\hbox{\rlap{\hbox{\lower4pt\hbox{$\sim$}}}\hbox{$>$}}}}
\def\v{\text{v}}

\title[Radio emission from colliding flows in V1723 Aql]{
Non-Thermal Radio Emission from Colliding Flows in Classical Nova V1723 Aql} 
\author[J. Weston et al.]{Jennifer H. S. Weston,$^1$
  J. L. Sokoloski,$^1$ Brian D. Metzger,$^1$ Yong Zheng,$^1$ Laura Chomiuk,$^{2}$
\newauthor  Miriam I. Krauss,$^3$ Justin Linford,$^2$  Thomas Nelson,$^4$  Amy Mioduszewski,$^3$   Michael P. Rupen,$^{5}$
\newauthor  Tom Finzell,$^2$and Koji Mukai.$^{6,7}$ \\
$^{1}$Columbia Astrophysics Laboratory, Columbia University, New York,
NY 10027, USA\\ 
$^{2}$Department of Physics and Astronomy, Michigan State University,
	East Lansing, MI 48824, USA\\
$^{3}$National Radio Astronomy Observatory, P.O. Box O, Socorro, NM 87801, USA\\
$^{4}$School of Physics and Astronomy, University of Minnesota,
	116 Church Street SE, Minneapolis, MN 55455, USA\\
$^{5}$ National Research Council of Canada, Herzberg Astronomy and Astrophysics Programs, Dominion Radio Astrophysical Observatory,\\ P.O. Box 248, Penticton, BC V2A 6J9, Canada\\
$^{6}$CRESST and X-ray Astrophysics Laboratory, NASA/GSFC, Greenbelt,
	MD 20771, USA\\
$^{7}$Department of Physics, University of Maryland, Baltimore County,
	1000 Hilltop Circle, Baltimore, MD 21250, USA}

\begin{document}

\maketitle
\label{firstpage}

\begin{abstract}
\noindent

The importance of shocks in nova explosions has been highlighted by {{\it Fermi}'s} discovery of $\gamma$-ray producing novae{.} {Over} three years of multi-band {VLA} radio observations of {the 2010} nova {V1723~Aql show} that shocks between fast and slow flows within the ejecta led to the acceleration of particles and the production of synchrotron radiation. {Soon} after the start of {the eruption}, shocks in the ejecta produced an unexpected radio flare, {resulting} in a {multi-peaked radio light curve}. {The emission eventually became consistent with} an expanding thermal remnant with {mass} $2 \times 10^{-4} M_\odot$ and {temperature}  $10^4$~K. {However, during the first two months, the $\gtrsim$10$^6$~K brightness temperature} at low {frequencies was} too high to be due to thermal emission from the small amount of X-ray producing shock-heated gas. Radio {imaging showed structures} with velocities {of}  $400$~km~s$^{-1}\,({\rm d}/6\,{\rm kpc})$ in the plane of the sky, perpendicular to a more {elongated} $1500$~km~s$^{-1}\, ({\rm d}/6\,{\rm kpc})$ {flow}. The morpho-kinematic structure of the ejecta from V1723~Aql {appears similar to} nova V959~Mon, {where} collisions between a {slow torus} and a faster {flow collimated} the fast flow and gave rise to $\gamma$-ray producing shocks. {Optical} spectroscopy and X-ray observations of V1723~Aql {during} the radio flare are consistent with this picture. {Our observations} support the idea that shocks in novae occur when a fast flow collides with {a slow} collimating torus. Such shocks could be responsible for hard X-ray emission, $\gamma$-ray production, and double-peaked radio light curves from some classical novae.

\end{abstract}


\section{Introduction}
\indent

The most common type of stellar explosion, a nova occurs when
accretion of material onto a white dwarf (WD) from its binary
companion ignites a thermonuclear runaway, expelling matter and
causing the system to brighten. Because novae occur frequently  -- an
estimated 35 yr$^{-1}$ in our galaxy alone \citep{Shafter97} -- they
can be used to probe stellar explosions, astrophysical outflows,
nuclear burning, and accretion. However, to find an accurate model of
these eruptions, one must {account} for 
the presence of shocks, bipolar outflows, rings, clumping, and other
asymmetries that have been shown to be present in different nova
systems \citep[e.g][]{Slavin95, Shara97, Sokoloski08, Chomiuk14}.  

Radio observations of novae furnish a useful tool for recording the evolution of a nova's expanding shell and understanding the properties of the ejecta. In the radio, the 
{ejecta remain optically thick to thermal emission} at significantly lower densities than at shorter wavelengths. The ejecta can therefore typically be detected for months to years in the radio. In classical novae, radio emission is generally observed to be dominated by thermal bremsstrahlung radiation, interpreted as emanating from an expanding shell \citep{Hjellming79, Seaquist77, Bode08}.

While non-thermal emission has long been evident in nova systems with
{red-giant companions}, such as RS Oph \citep{Taylor89}, in classical
novae with main sequence companions, the dominant features in
historical radio light curves and spectra are generally consistent
with an expanding thermal shell producing free-free emission
\citep{Hjellming79, Seaquist80, Hjellming96, Bode08}. Given some
reasonable assumptions about the temperature and morphology of a
thermal shell, comprehensive radio spectral coverage as the ejecta
transition from being optically thick to optically thin enables
observers to infer the density profile of the nova shell and estimate
the total mass ejected. Resolved images in radio allow for further
constraints on the morphology, temperature, and evolution of the
thermal shell. However, only a handful of classical novae have been
well observed at multiple radio frequencies 
from the optically thick rise to the optically thin decay, including HR Del \citep{Hjellming79}, V1500 Cyg \citep{Seaquist80}, and V1974 Cyg \citep{Hjellming96}. This lack is particularly noticeable in the first few months immediately following the outburst, when the thermal radio emission is expected to be faint. 

Technological improvements to the Karl G. Jansky Very Large Array (VLA) and other radio telescopes within the past five years have vastly increased the sensitivity of radio observations. Our group has been using the updated capabilities of the VLA in concert with optical and X-ray observations to create a more complete and multi-wavelength picture of the mass-loss history of novae. V1723~Aql was the first classical nova to be discovered in outburst after the upgrade to that facility. Beginning two weeks after the initial discovery of the eruption of V1723~Aql in 2010 September, the observations of \citet{Krauss11}, in combination with the additional radio monitoring presented here, have produced one of the best, most detailed radio light curves of a classical nova to date.

Discovered in outburst at {its} optical maximum (mag 12.4) on 2010
September 11, and with a {magnitude of fainter than} 14.0 on 2010 September
10, we take that discovery of V1723~Aql as the start of the outburst,
$t_0=$ MJD 55450.5 \citep{Yamanaka10, Balam10}. Optical observations
revealed V1723~Aql to be a typical `fast nova' \citep{Payne57} of the
Fe II spectral type, fading by two magnitudes within 20 days
\citep{Yamanaka10}. Despite the lack of extensive optical or IR
followup, the observations in the first few weeks of the erruption showed rapid reddening, implying the formation of dust particles as early as 20 days after the start of the outburst \citep{Nagashima13}. Emission lines of H$\alpha$ on the day of discovery had full width at zero intensity of 3000 km s$^{-1}$, implying a maximum expansion velocity $\v_{\rm obs}=$1500 km s$^{-1}$ \citep{Yamanaka10}. The nova was neither embedded in the wind of an M-giant mass donor nor a recurrent nova -- making it representative of the majority of novae \citep{Nagashima13}. 

Although V1723~Aql showed no unusual aspects in its early optical
behaviour, the VLA observations 
{revealed} an unexpected radio flare peaking roughly 45 days after the start of the optical outburst \citep{Krauss11}. \citet{Krauss11} found that during this flare, the radio flux density rose as t$^{3.3}$, far more rapidly than the t$^2$ expected {from freely} expanding, isothermal, optically thick thermal ejecta. The radio spectrum during the flare was also inconsistent with that of an optically thick thermal source of a fixed temperature \citep{Krauss11}. With only 175 days of radio monitoring, \citet{Krauss11} were not able to conclusively determine the origin of this flare, or {whether
V1723~Aql} would evolve {according to} 
the more standard picture of radio emission from a classical nova after this unexpected activity. We label the observations between the start of the eruption and day 175, presented in \citet{Krauss11}, as the {\em early-time} observations, and the new observations presented here (between day 198 and day 1281) as the {\em late-time} observations.

We describe below how the combination of the {\em early-time} and {\em
  late-time} radio observations {reveals} that whereas the late-time
observations are dominated by the expected thermal emission {from}
a freely expanding nova shell, there is strong evidence that the
early-time flare was dominated by
non-thermal emission.
In section 2 we present our late-time observations and data reduction
procedure; in section 3, we describe the observational findings and
data analysis; {in} section 4, {we introduce} a thermal model 
for {the remnant} after day 200, {explore} the possibility of
the early-time emission being produced by relativistic particles
accelerated in internal shocks, and 
{propose} that the ejecta from V1723~Aql had a morpho-kinetic structure
with internal shocks like {that} of $\gamma$-ray producing nova
V959~Mon. In section 5, we highlight our conclusions. 

\section{Observations and Data Processing}

\begin{table*}
 \centering

 \begin{minipage}{138mm}
 \caption{VLA observations of V1723~Aql, between 2010 September and 2014 March.}
  \label{tab:ObsLog}
  \begin{tabular}{@{}lccccclc@{}}
  \hline
   Observation Date     & MJD  & Day$^a$ & Epoch     & VLA Program ID &Configuration&Observed Bands& Time$^b$(min)    \\

  \hline
   \multicolumn{6}{l}{{\bf Early time observations}  \citep[previously presented in][]{Krauss11}}\\
  \hline
2010 September 25 	& 55464.1	&	13.6		& 0 & 10B-200 & DnC 	& C, Ka & 57 \\ 
2010 October 3 	& 55472.1	&	21.6		& 1 & 10B-200 & DnC 	& C, Ka & 21 \\ 
2010 October 6 	& 55475.1	&	24.6		& 2 & 10B-200 & DnC/C 	& C, Ka & 58 \\ 
2010 October 15	& 55483.1	&	32.6		& 3 & 10B-200 & C 		& C, Ka & 48 \\ 
2010 October 18 	& 55487.1	&	36.6		& 4 & 10B-200 & C 		& L, C, X & 43 \\ 
2010 October 19 	& 55488.1	&	37.6		& 4 & 10B-200 & C 		& C, Ka & 28 \\ 
2010 October 24 	& 55493.1	&	42.6		& 5 & 10B-200 & C 		& L, C, X & 43 \\ 
2010 October 29 & 55497.0	&	46.5		& 6 & 10B-200 & C 		& L, C, X, Ka & 71 \\ 
2010 November 7	& 55507.8	&	57.2		 & 7 & 10B-200 & C		 & L, C, X, Ka & 74 \\ 
2010 December 3 	& 55533.9	&	83.4		& 8 & 10B-200 & C 		& L, C, X, Ka & 84 \\ 
2010 December 18 	& 55548.9	&	98.4		& 9 & 10B-200 & C 		& L, C, X & 52 \\ 
2010 December 21 	& 55551.7	&	101.2	& 9 & 10B-200 & C 		& C, Ka & 34 \\ 
2011 January 14 	& 55575.8	&	125.3	& 10 & 10B-200 & C 		& L,C,X, Ka & 70 \\ 
2011 January 30 	& 55591.8	&	141.3	& 11 & 10B-200 & CnB 	& L,C,X, Ka & 63 \\ 
2011 March 5	& 55625.5	&	175.0	& 12 & 11A-254 & B 		& C, X, Ka & 51 \\   
  \hline
 \multicolumn{6}{l}{\bf Late time observations} \\
  \hline

2011 March 28 	& 55648.6 & 198.1 	& 13 & 11A-254 & B & C, X, Ka & 52 \\ 
2011 August 2 		& 55775.2 & 324.7 	& 14 & 11A-254 & A & C, X, Ku, K, Ka & 116 \\ 
2011 August 27 	& 55800.2 & 349.7	& 15 & 11A-254 & A & C, X, Ku, K, Ka & 72 \\ 
2011 November 11 	& 55876.1 & 425.6	& 16 & 11B-170 & D & C, X, Ku, K, Ka & 58 \\ 
2012 January 14 	& 55940.8 & 490.3	& 17 & 11B-170 & DnC & C, X, Ku, K, Ka & 58 \\ 
2012 April 19 		& 56037.0 & 586.5	& 18 & 11B-170 & C & C, X, Ku, K, Ka & 58 \\ 
2012 December 8 	& 56269.0 & 818.5	& 19 & 12B-226 & A & C, X, Ku, K & 20 \\ 
2012 December 22 	& 56283.6 & 832.5 	& 20 & 12B-226 & A & X, Ka & 45 \\ 
2013 March 27 	& 56378.4 & 927.9	& 21 & 13A-465 & D & C, X, Ku, K, Ka & 106 \\ 
2013 August 30 	& 56534.1 & 1083.6	& 22 & 13A-461 & C & L, C, X, Ku, K, Ka & 98 \\ 
2013 November 30 	& 56626.7 & 1176.5	& 23 & 13A-465 & B & L, S, X, Ka & 101 \\ 
2014 March 15		& 56731.5 & 1281.0 	& 24 & 13A-465 & A & X, Ku & 112 \\  
\hline
\end{tabular}\\
$^a$ Days after initial detection of outburst, 2010 September 11. \\
$^b$ Total time of observations on-source. 
\end{minipage}
\end{table*}

\indent

Building on the work of \citet{Krauss11}, who used VLA observations
during the first six months of the outburst of V1723~Aql to identify
the bright early-time flare, we present our observations of V1723~Aql
with the VLA between 2011 March (day 198) and 2014 March (day 1281) --
from six months to three and a half years after the start of the
eruption (see Table~\ref{tab:ObsLog}). For completeness,
Table~\ref{tab:ObsLog} additionally lists the early-time observations
of \citet{Krauss11}. The late-time observations were taken using the
VLA WIDAR correlator at a range of frequencies spanning 2.5-36.5~GHz
with bandwidths of 2048~MHz, distributed in two tuneable 1048-MHz
sidebands and using 8-bit samplers. The central frequencies in each
band may have had some slight variation due to frequency dependent flagging in post-processing to minimize radio frequency interference (RFI). At each frequency, on-source observations were alternated with observations of a nearby phase reference calibrator (J1851+0035, J1832-1035, or J1822-0938), with an additional observation of a standard flux calibration source (either 3c286 or 3c48). We used observations of the unpolarized calibrator J0319+4130 (3c84) for polarization leakage solutions. The late-time data presented here comprise 29.5~hours of observations (14.9 hours on source V1723~Aql) and 658~GB of raw data over three years and two full cycles of VLA configurations. Including the earlier programs presented in \citet{Krauss11}, the data total 1604~GB and 59.5 hours (28.2 hours on source) between 2010 September 25 and 2014 March 15.

\begin{figure*}
\centering 
\begin{minipage}{180 mm}
\caption{Observed flux densities of V1723~Aql between 2010 September and 2014 March. Error bars are as reported in Table~\ref{tab:Flux}, but may be too small to be visible.}
\includegraphics[width=180 mm]{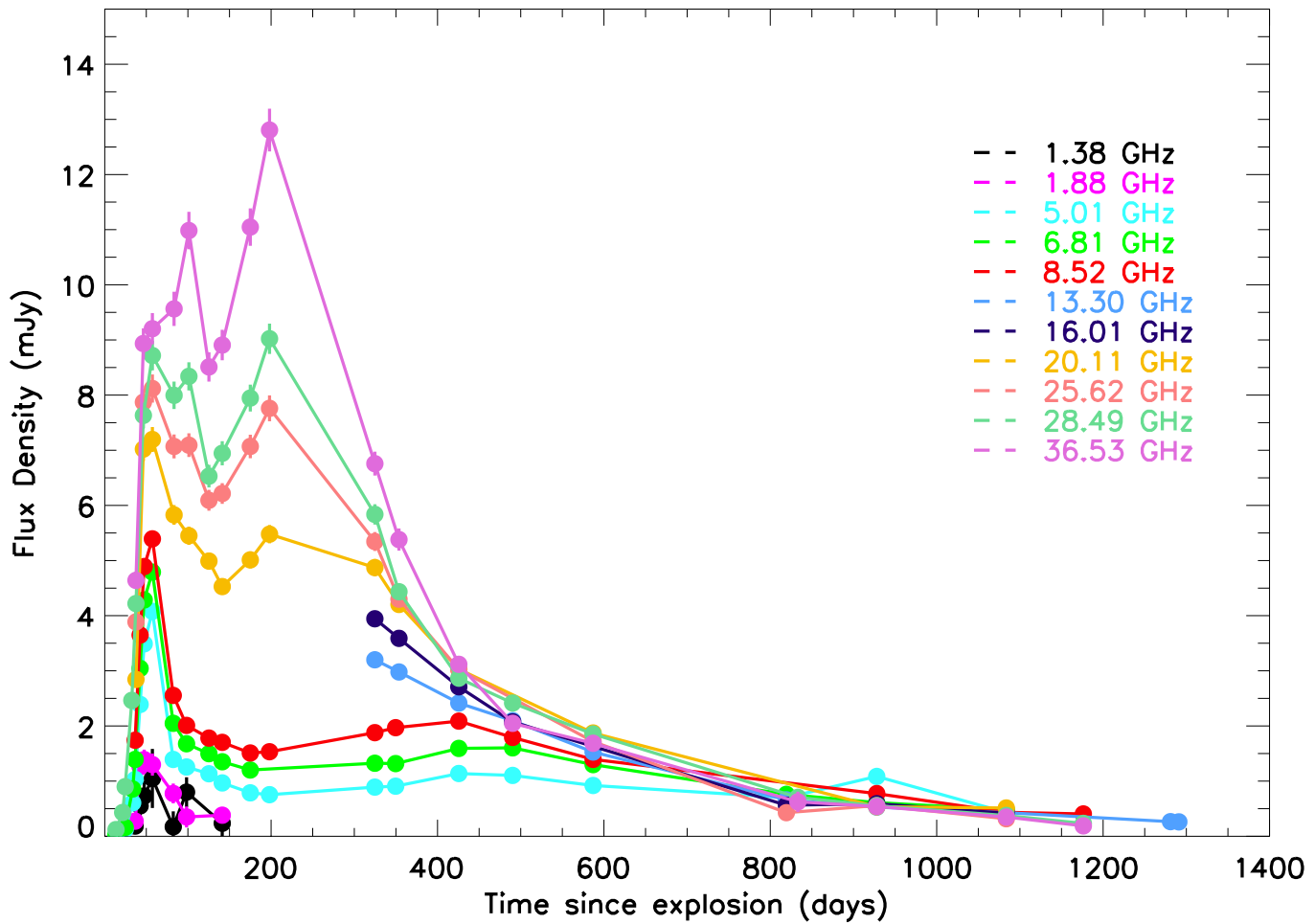}
\label{fig:FullFlux} 
\end{minipage}
\end{figure*}

\begin{figure*}
\centering
\begin{minipage}{180 mm}
\caption{Spectra of V1723~Aql between 2011 March and 2014 March, with lines 
{showing the best-fit spectral indices} $\alpha$ (where $S_\nu \propto \nu^\alpha$). For epochs where the spectral index appears to be a function of frequency, 
{spectral indices are listed}
from lowest frequency range to highest. See Table~\ref{tab:SpecIndex}}
\includegraphics[width=180 mm]{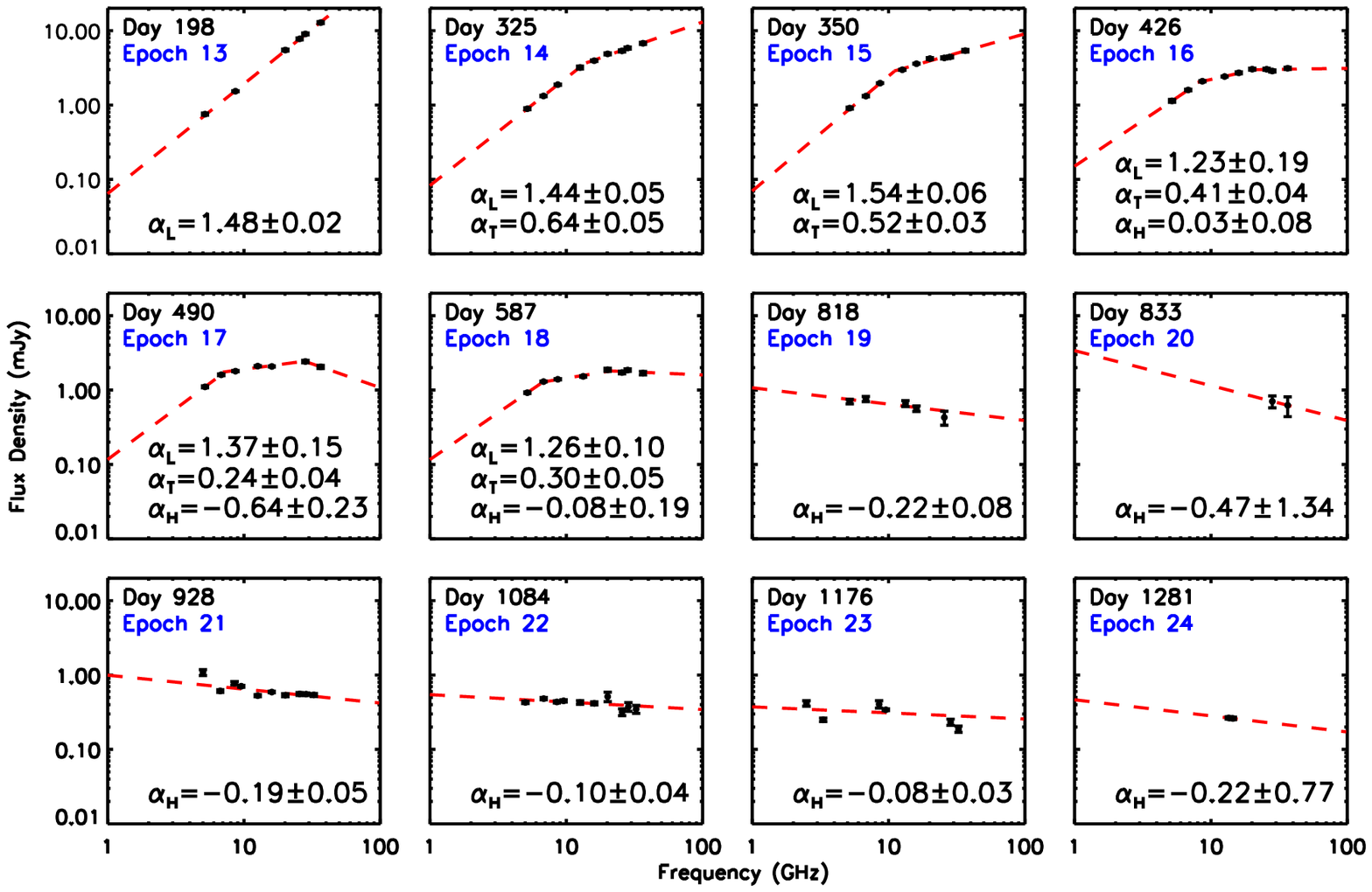}
\label{fig:LateSpec} 
\end{minipage}
\end{figure*}

We processed all data using NRAO's Common Astronomical Processing
System (CASA) or Astronomical Image Processing System (AIPS). As is
standard procedure, we obtained a bandpass solution and used it to
solve for gain amplitude and phase, then applied these solutions to
the source, using the `Perley-Butler 2010' flux density scale. Imaging
was completed using the CLEAN algorithm \citep{Hogbom74} with Briggs
weighting, with the additional use of superresolution techniques
\citep{Staveley93} on day 1281 to resolve small structures during that
observation. Radio flux densities were measured by fitting two
dimensional gaussians to the emitting region in the image plane with
the task {\it imfit}. We estimated uncertainties in flux density by
adding in quadrature the error from the gaussian fit and systematic
errors of 1\% at frequencies below 19~GHz and 3\% at those above, with
the results presented in Table~\ref{tab:Flux} and
Figs~\ref{fig:FullFlux} and \ref{fig:LateSpec}. 
{All error bars correspond to 1}$\sigma$ {uncertainties.}

\section{Data Analysis and Results}

\begin{table*}
  \caption{Flux densities for late-time observations of V1723~Aql }
 \begin{minipage}{180mm}
  \label{tab:Flux}
  \begin{tabular}{@{}cccccccc@{}}
  \hline
{\bf Epoch} & {\bf Day}  & \multicolumn{6}{c}{\bf Observed flux density (mJy)}\\
& & \multicolumn{1}{r|}{}{\it 2.5 GHz (S band)} & \multicolumn{1}{r|}{\it 3.3 GHz (S band)} & \multicolumn{1}{r|}{\it 5.2 GHz (C band)} & \multicolumn{1}{r|}{\it 6.8 GHz (C band)} & \multicolumn{1}{r|}{\it 8.7 GHz (X band)} &\multicolumn{1}{r|}{\it 9.5 GHz (X band)} \\ \hline
13&198 &  ...* & ... & 0.75 $\pm$  0.02 & ... & 1.53 $\pm$ 0.03 & ...      \\ 
 14&325 & ... & ... & 0.89 $\pm$  0.03 & 1.32 $\pm$ 0.03 & 1.88 $\pm$ 0.03& ...    \\ 
15&250 & ... & ... & 0.91 $\pm$ 0.03 & 1.33 $\pm$ 0.03 & 1.97 $\pm$ 0.03 &  ...  \\ 
16&425 & ... & ... & 1.14 $\pm$ 0.05 & 1.59 $\pm$ 0.04 & 2.09 $\pm$ 0.04 & ...   \\ 
 17&490 & ... & ... & 1.10 $\pm$ 0.03 & 1.60 $\pm$ 0.05 & 1.79 $\pm$ 0.03&  ...   \\ 
 18&587 & ... & ... & 0.92 $\pm$  0.02 & 1.30 $\pm$ 0.02 & 1.39 $\pm$ 0.03 &  ...  \\ 
 19&818 & ... & ... & 0.70 $\pm$ 0.05  & 0.76 $\pm$ 0.06 & ... 				& ...   \\ 
 21&928 & ... & ... & 1.09 $\pm$ 0.10 & 0.61 $\pm$ 0.03 & 0.77 $\pm$ 0.04& 0.71 $\pm$ 0.02    \\ 
 22&1084 & ... &	 ... & 0.43 $\pm$ 0.02 &  0.48 $\pm$ 0.02 &	0.44 $\pm$ 0.02 &	0.45 $\pm$ 0.02 \\
 23&1177 & 0.42 $\pm$ 0.04 & 0.25 $\pm$ 0.02 & ... & ... & 0.41 $\pm$ 0.05&   0.34 $\pm$ 0.01   \\ 
\hline
{\bf Epoch} & {\bf Day}  & \multicolumn{6}{c}{\bf Observed flux density (mJy)}\\
 & & \multicolumn{1}{r|}{\it 12.9 GHz (Ku band)} & \multicolumn{1}{r|}{\it 16.1 GHz (Ku band)} & \multicolumn{1}{r|}{\it 20.1 GHz (K band)} & \multicolumn{1}{r|}{\it 25.6 GHz (K band)} & \multicolumn{1}{r|}{\it 28.5 GHz (Ka band)} & \multicolumn{1}{r|}{\it 36.5 GHz (Ka band)} \\ \hline
13& 198 &  ... & ...  & 5.48 $\pm$ 0.17   & 7.76 $\pm$ 0.23 & 9.02 $\pm$ 0.28 & 12.81 $\pm$ 0.39 \\ 
14 & 325 &  3.20 $\pm$ 0.12    & 3.94 $\pm$ 0.10  & 4.87 $\pm$ 0.16   & 5.35 $\pm$ 0.17 & 5.84 $\pm$ 0.18 & 6.76 $\pm$ 0.22 \\ 
15 & 350 &  2.98 $\pm$ 0.05   & 3.59 $\pm$ 0.05 & 4.20 $\pm$ 0.13   & 4.30 $\pm$ 0.14 & 4.43 $\pm$ 0.14 & 5.38 $\pm$ 0.20 \\ 
16 & 426 & 2.42 $\pm$ 0.03   & 2.71 $\pm$ 0.08   & 3.03 $\pm$ 0.10   & 3.03 $\pm$ 0.10 & 2.87 $\pm$ 0.09 & 3.12 $\pm$ 0.11 \\ 
17 & 490 &  2.10 $\pm$ 0.03   & 2.07 $\pm$ 0.03 & ...   & ...   & 2.42 $\pm$ 0.09 & 2.05 $\pm$ 0.10 \\ 
18 & 587 &  1.53 $\pm$ 0.04   & ... & 1.87 $\pm$ 0.08   & 1.72 $\pm$ 0.07 & 1.85 $\pm$ 0.08 & 1.69 $\pm$ 0.10 \\ 
19 & 818 &  0.66 $\pm$ 0.06   & 0.56 $\pm$ 0.05  & ... & 0.43 $\pm$ 0.09 & ... & ... \\ 
20 & 832 & ... & ... & ... & ... & 0.71 $\pm$ 0.13 & 0.62 $\pm$ 0.19 \\ 
21 & 928 &  0.53 $\pm$ 0.02  & 0.59 $\pm$ 0.02    & 0.54 $\pm $ 0.03 & 0.56 $\pm$ 0.03 & 0.55 $\pm$ 0.02 & 0.54 $\pm$ 0.03$^c$ \\ 
22 & 1084 & 0.43 $\pm$ 0.03   &	0.42 $\pm$ 0.02 &	0.51 $\pm$ 0.08 & 0.32 $\pm$ 0.03 & 0.38 $\pm$ 0.05& 0.35 $\pm$ 0.04$^c$\\
23 & 1177  & ... & ... & ... & ... & 0.23 $\pm$ 0.02 & 0.19 $\pm$ 0.02$^c$ \\ 
24 &  1281 &  0.27 $\pm$ 0.01$^a$ & 0.26 $\pm$ 0.02$^b$    & ... & ... & ... & ... \\ 
\hline
\end{tabular}
* `...' indicates no observation at this frequency on this day. \\
$^a$ Values are reported for 13.5 GHz.\\
$^b$ Values are reported for 14.5 GHz.\\
$^c$ Values are reported for 32.5 GHz.
\end{minipage}
\end{table*}

\begin{table*}
 \centering
 \begin{minipage}{144mm}
    \caption{Spectral indices. For days where we fit models with
      spectral breaks, $\nu_{\rm lt}$  refers to the break between the
      low-frequency and transition parts of the spectrum, and
      $\nu_{\rm th}$ refers to the break between the transition and
      high-frequency portions of the spectrum. $\alpha_{\rm low}$,
      $\alpha_{\rm trans}$, and $\alpha_{\rm h}$ are the corresponding
      spectral indices for these segments. $\beta$ shows the density
      profile $n \propto 1/r^\beta$ associated with the spectral index
      $\alpha$ for a spherical shell transitioning between optically
      thick and optically thin thermal emission \citep{Wright75}, and
      $\chi^2_{r}$ and $d\!f$ are the reduced chi-square 
statistic and the number of degrees of freedom, respectively.} 
  \label{tab:SpecIndex}
\begin{tabular}{@{}ccccccccccc@{}}
  \hline

{\bf  Epoch} & {\bf Day} & {\bf model}* 	& $\alpha_{low}$ 		& $\alpha_{trans}$	& $\alpha_{h}$ 		& $\nu_{lt}$  ({\rm GHz})	& $\nu_{th}$ ({\rm GHz}) & $\chi^2_{r} $ &	$d\!f$		& $\beta$	 			\\ \hline
 13 	 &  198	& SPL 	& 1.48 $\pm$ 0.02	& 				& 				& 					   	&					&		0.73		& 4	&					\\	
 14 	 &  325	& BPL 	& 1.44 $\pm$ 0.05	&  0.64 $\pm$ 0.05	& 				& 		13.6 			&					&		0.45		& 7	&	2.04	$\pm$ 0.06	\\	
 15 	  & 	350 & BPL 	& 1.54 $\pm$ 0.06	&  0.52 $\pm$ 0.03	& 				& 		11.3 	 		&					&		2.75		& 7	&	1.92	$\pm$ 0.03	\\	
 16 	  & 426 	& DBPL 	& 1.23 $\pm$ 0.19	&  0.41 $\pm$ 0.04	& 0.03 $\pm$ 0.08	& 		8.3 	 		&		21.2 		&		0.60		& 7	&	1.84 $\pm$ 0.03	\\	
 17 	  & 490	& DBPL 	& 1.37 $\pm$ 0.15	&  0.24 $\pm$ 0.04	& -0.64 $\pm$ 0.23	& 		7.3 	 		&		27.8 		&		2.53		& 5	&	1.69 $\pm$ 0.03	\\	
 18 	  & 587	& DBPL 	& 1.26 $\pm$ 0.10	&  0.30 $\pm$ 0.05	& -0.08 $\pm$ 0.19 	& 		6.7 	 		&		21.2 		&		1.00		& 6	&	1.74 $\pm$ 0.04	\\	
19 	  & 818	& SPL 	& 				& 				& -0.22 $\pm$ 0.08	& 					   	&					&		1.26		& 3	&					\\	
20 	  & 832	& SPL 	&  				& 				& -0.47 $\pm$ 1.34	& 					   	&					&		--		& 0	&					\\	
21 	&  928 	& SPL 	&  				& 				& -0.19 $\pm$ 0.02	& 					   	&					&		8.40		& 8	&					\\	
22 	 &  1084	& SPL 	&  				& 				& -0.10 $\pm$ 0.04	& 					   	&					&		2.12		& 8	&					\\	
23 	   & 1177	& SPL 	&  				& 				& -0.08 $\pm$ 0.03 	& 					   	&					&		24.19	& 4	&					\\	
24 	  & 	1281& SPL 	&  				& 				& -0.22 $\pm$ 0.77	& 					   	&					&		--		& 0	&					\\	

\hline
\end{tabular}
*SPL -- single power law fit\\
BPL -- broken power law fit (one break)\\
DBPL -- double broken power law fit (two breaks)
\end{minipage}
\end{table*}

\subsection{Flux density and spectral evolution}
\indent

Around day 45, the flux densities reached a local maximum and
subsequently declined at all frequencies, with another local maximum
at all frequencies between days 200 and 400 (see Table~\ref{tab:Flux}
and Fig.~\ref{fig:FullFlux}). We refer to the radio brightening that
peaked around day 45 as the {\em early-time flare}.  There was an
additional local maximum between these two bright peaks at the higher
frequencies. The flux densities at the highest frequencies reached
higher values than those at lower frequencies during both major
peaks. For example, during the final local maxima (between days 200
and 400), the 36.5~GHz flux density peaked at $12.80 \pm 0.39$~mJy, whereas the 5.2~GHz flux density peaked at $1.14 \pm 0.05$~mJy. Moreover, the flux densities at {\it Ka}-band (28.5/36.5~GHz) reached their highest values on day 198, whereas the flux densities at lower frequencies peaked later, with {\it C}-band flux densities (5.2/6.8~GHz) peaking on day 426 or 490.

During the {late-time 
period}, the radio spectrum evolved from one that rose with frequency to one that fell slightly with frequency (see Fig.~\ref{fig:LateSpec} and Table~\ref{tab:SpecIndex}). To obtain spectral indices and 1$\sigma$ uncertainty estimates for those spectral indices for each observation ($S_\nu \propto \nu^\alpha$, where $S_\nu$ is the flux density at frequency $\nu$, and $\alpha$ is the spectral index), we fit data with a power-law using IDL's `linfit' function, which minimizes the $\chi^2$ statistic, with the data and error bars in log space. With the simple single power-law model, we found acceptable fits ($\chi^2_{r} \lesssim $1.5, where $\chi^2_{r}$ is the reduced $\chi^2$ value) only for days 198 and 818, between which time the spectral index fell from $\alpha=1.47 \pm 0.05$ to $\alpha=-0.22 \pm 0.08$. 

The late-time transition from a rising to a falling spectrum occurred
via the appearances of first one, and then a second, spectral
{break}. We refer to the three distinct portions of the spectrum when
two breaks were present as the {\em low-frequency}, {\em transition},
and {\em high-frequency} parts of the spectrum. We refer to the
corresponding spectral indices as $\alpha_{\rm low}$, $\alpha_{\rm
  trans}$, and $\alpha_{\rm h}$, and the frequencies at which the
breaks between the regions of different spectral index occur as
$\nu_{\rm lt}$ (for the break between the low-frequency and transition
part of the spectrum) and $\nu_{\rm th}$ (for the break between the
transition and high-frequency portion of the spectrum). For epochs
where both a poor $\chi^2$ fit and the residuals of the simple
power-law model suggested a spectral break, we fit the data with
broken power law models with one or two
breaks. Fig.~\ref{fig:LateSpec} shows the spectra and fits, and
Table~\ref{tab:SpecIndex} reports the power law models that combined to minimize $\chi^2$, with the break {frequencies indicating where the power-law models intersect.
The uncertainties for the break frequencies correspond to the range of} frequencies at which the power laws intersected given a $\pm 1\sigma$ range for the spectral indices. We were able to find acceptable fits for a single break power-law model on day 325, and for a double-break power law model on days 426 and 587. 

Fig.~\ref{fig:LateSpec} and Table~\ref{tab:SpecIndex} show {the evolution of the spectrum and its component segments.}
The low-frequency spectral index $\alpha_{\rm low}$ typically had values of approximately 1.4, $\alpha_{\rm trans}$ typically had values between 0.2 and 0.6, and $\alpha_{\rm h}$ typically had values of approximately $-0.2$. The spectrum on day 490 (see Fig.~\ref{fig:LateSpec}) shows the characteristic low-frequency, transition, and high-frequency spectral regions most clearly. On day 198, the spectrum was well described by a simple power-law with a low-frequency type spectral index. On day 325, the lowest frequencies maintained a steep rising spectrum of $\alpha_{\rm low} = 1.44 \pm 0.05$; above 16.1~GHz the spectrum became shallower, with $\alpha_{\rm trans} = 0.64 \pm 0.05$. The {spectra continued to show this break} between $\alpha_{\rm  low}$ and $\alpha_{\rm trans}$ in observations between days 325 and 587, during which time $\nu_{\rm lt}$ decreased from 13.6~GHz to 6.7~GHz. Observations between days 426 and 587 showed the additional break at higher frequencies, $\nu_{\rm th}$, above which the spectrum flattened to $\alpha_{\rm h}=0.13 \pm 0.13$ on day 426, $\alpha_{\rm h}=-0.64 \pm 0.23$ on day 490, and $\alpha_{\rm h}=-0.08 \pm 0.19$ on day 587. While the very small uncertainties in our flux densities reveal some deviation from a single power law {during the last five observations, the spectrum was nevertheless approximately described by a fairly flat single power law during that time.}

\subsection{Resolved images}

\begin{figure}
\begin{minipage}{90 mm}
\caption{{\bf {Top panel:}}  Spatially resolved image of V1723~Aql taken on 2012 December 22 in {\it Ka}-band (32.4~GHz). When fitted to a two-dimensional gaussian, the brightness profile {has a FWHM along the} major axis of $160 \pm 10$~milliarcseconds (mas) and {along the} minor axis of $110 \pm 20$~mas, where the major axis has a position angle of $160^\circ \pm 10^\circ$ east of north. Contour levels are at [-2,2,3,4,5]$\times$RMS with RMS= $2.26 \times 10^{-5}$ Jy. There are two resolved components roughly perpendicular to the major axis of the gaussian profile, with a separation of 70~mas at an angle of $82^\circ$ E of N. 
{\bf {Bottom panel:}} Spatially resolved image of V1723~Aql taken on 2014 March 14 in {\it Ku}-band (14~GHz). When deconvolved with a beam size of 90 $\times$ 90~mas, we see a ring-like structure. Contour levels are at [-2, 2, 2$\sqrt{2}$, 4, $4\sqrt{2}$, 8, $8\sqrt{2}$, 16, $16\sqrt{2}$, 32]$\times$RMS with RMS=$7.01 \times 10^{-6}$ Jy.}
  \centering
\includegraphics[width=\columnwidth]{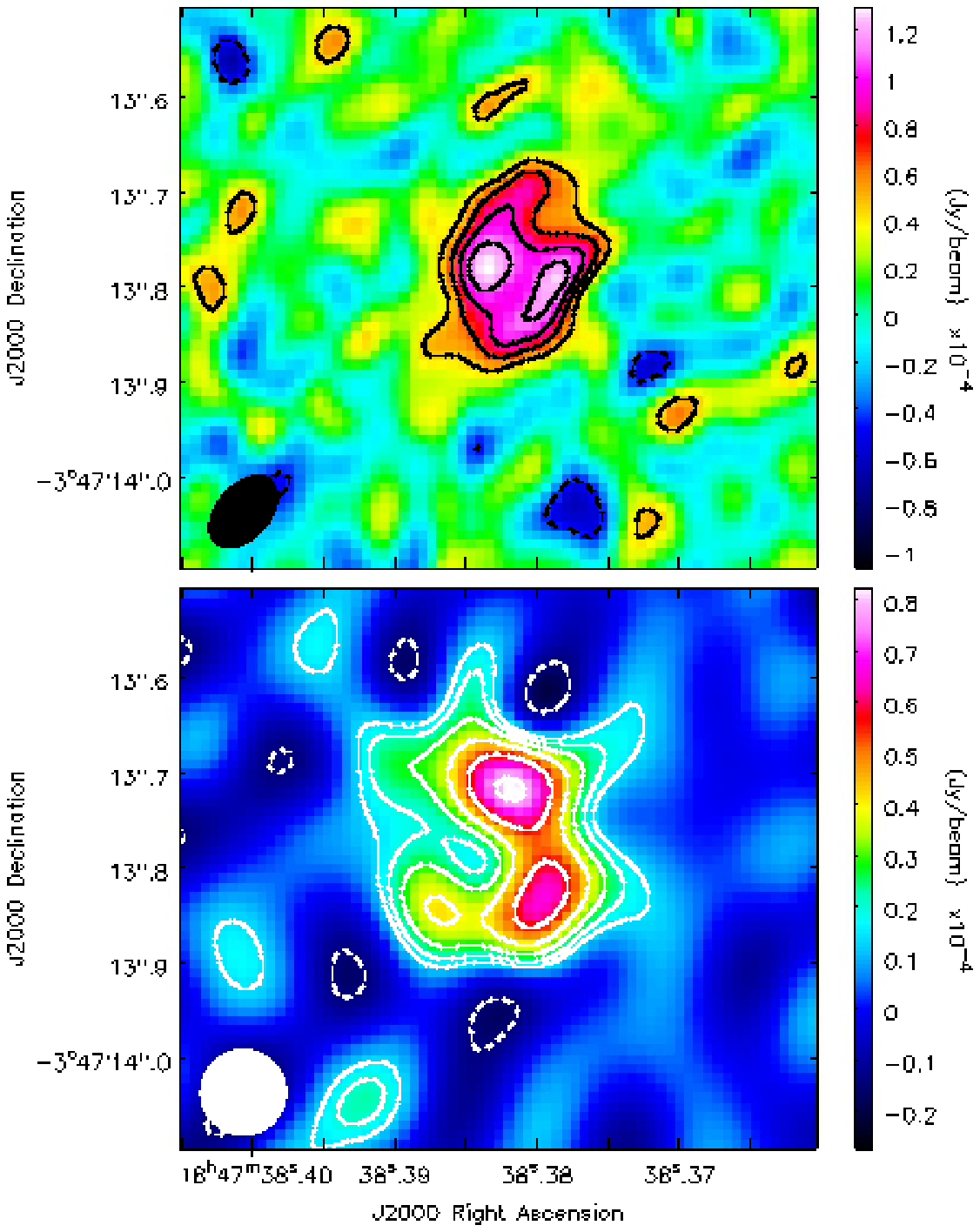}
 \label{fig:Resolved}
 \end{minipage}
 \end{figure}

\indent

V1723~Aql was resolved at the highest observed radio frequency on day
832 (Ka-band) and on day 1281 (Ku-band) (see Fig.~\ref{fig:Resolved},
panel a).  The remnant was therefore large enough to be spatially
resolved, with the
difference between the integrated and the peak flux density greater than
five times the root mean square error (RMS). 
{Fig.}~\ref{fig:Resolved} {shows images from}
the second and third of the three observations during
which the VLA was in A configuration (which gives the highest spatial
resolution). 
The image from day 832 (2012 December 22) at 32.4~GHz shows that the
radio emitting material was elongated in the NW/SE direction and had
visible substructure. The full width half maximum of the brightness
profile along the major axis of the fitted gaussian was $160 \pm
10$~milliarcseconds (mas), {with the FWHM along the minor axis of } $110
\pm 20$~mas. { The position angle (PA) of the major axis was}$160^\circ \pm
10^\circ$ (where PA is the east of north).  The clean beam size was 90~mas by 60~mas, and the peak
flux density was $130 \pm 20$~$\mu$Jy/beam, with an RMS of
22.6~$\mu$Jy/beam. 
To {more accurately} estimate the size scale of the source, we
fit the brightness profile, deconvolved from the synthesized clean
beam, to a {Gaussian.  Taking} the width of the Gaussian along the
major axis at a flux density {of twice the RMS as an indicator of size, the size scale} along the major axis on day 832 was $250 \pm 10$~mas. The 
substructure showed two resolved components {with} brightness
peaks separated by 70~mas.  {They were aligned perpendicular} to the major axis of the
elongated material, with a PA of about $80^\circ$. 

Taking into account the expected expansion by 50\% between day 832 and
day 1281, the size of the radio-emitting remnant on day 1281 indicates
that the material that was most extended on day 832 had faded, and
only the {denser} inner structure remained detectable.
By day 1281 (2014 March 14; using robust $= 0.5$ in the CLEAN algorithm),
{the resolved material} was elongated in the NE/SW direction, instead of
the NW/SE {elongation} seen on day 832. The fitted gaussian had a 
{FWHM} of $180 \pm 10$~mas {along the major axis (which had a PA of} $35^\circ \pm 5^\circ$), and a {FWHM along the minor} 
axis of $133 \pm 7$~mas. 
The peak flux density was $118.5 \pm
3.5$~$\mu$Jy/beam, with RMS of 3.32~$\mu$Jy/beam, resulting in a size
scale of $408 \pm 2$~mas. With a clean beam size of 160~mas by
130~mas, no substructure was resolved. If we consider the two
components visible on day 832, and extrapolate from their 70~mas
separation, we would have expected a separation of 108~mas on day 1281
-- which was not within the formal resolution limits of this
observation. Therefore, we additionally used superresolution
techniques \citep{Staveley93} to examine the more detailed structure
of the source on this day, deconvolving the source with the slightly
smaller beam size of 90 by 90~mas. Using this technique, we uncovered
a ring-like structure with dimensions $\sim 0.22 \times 1.5$~arcsec
and PA 45$^\circ$ (see Fig.~\ref{fig:Resolved}, panel b), with an RMS
of 7.0~$\mu$Jy/beam. 

\section{Discussion}
\begin{figure}
\begin{minipage}{\columnwidth}
\centering
\caption{We propose a picture where a fast, bipolar outflow is shaped by {a slow} moving equatorial torus. Shocks arose in the collision region between the slow and fast flows.}
\includegraphics[width=\columnwidth]{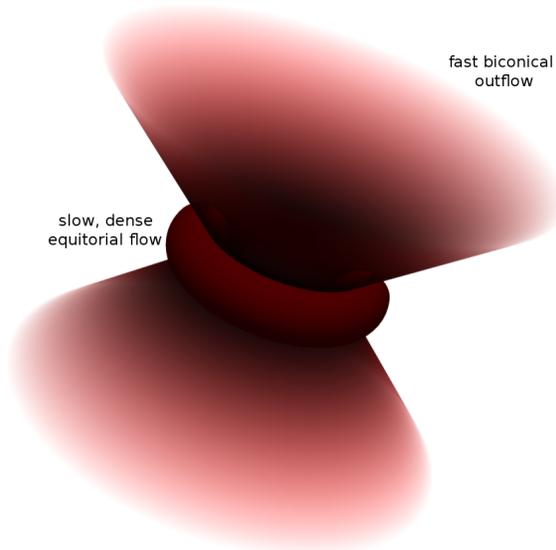}
\label{fig:Cartoon} 
\end{minipage}
\end{figure}
\indent

We propose that the radio emission from V1723~Aql -- when considered in the context of published optical and X-ray observations -- is most consistent with a picture in which the nova initially generated a slow outflow and then a faster flow. We argue below that the slow flow had a speed of a few hundred km~s$^{-1}$, and that it was likely concentrated toward the equatorial plane of the binary. The faster flow had a maximum speed of at least 1500 km~s$^{-1}$, and it collided with the early, slow ejecta within a few weeks of the start of the eruption, producing the early-time flare. The collisions led to shocks that heated some of the ejecta to temperatures above $10^7$~K, and also accelerated particles to relativistic speeds. The slow-moving material probably collimated the faster flow into a biconical structure. Fig.~\ref{fig:Cartoon} shows a cartoon of this toy model. The scenario that we propose for V1723~Aql is consistent with the one that \citet{Chomiuk14} suggested for the $\gamma$-ray producing nova V959~Mon.

In the discussion below, we present evidence that supports the picture of a fast flow that collided with, and was shaped by, a preceding slow outflow. To diagnose the flows without the complexity of non-thermal emission from the early-time shock, we begin in section {\em 4.1} by describing how the brightness temperature and radio spectral evolution at late times were consistent with a freely expanding thermal plasma with a temperature of around $10^4$~K. The density profile and radio images of this radio-emitting plasma suggest that it had a roughly bipolar or biconical outer structure, and a shell-like inner structure with a likely equatorial density enhancement. In section {\em 4.2}, we show that the early-time radio flare was probably dominated by synchrotron emission. Given the low X-ray flux during the flare \citep{Krauss11} and high brightness temperatures, it cannot be attributed to thermal emission alone. In section {\em 4.3}, we discuss how V1723~Aql relates to other classical novae in its behaviour and shape, including those that produce $\gamma$-rays.

\subsection{Origin of the Late-Time Radio Emission: Expanding thermal shell}
\subsubsection{General physical constraints}
\begin{table*}
 \centering
 \begin{minipage}{123 mm}
  \caption{  Total ejecta mass M and minimum velocity ($v_i$) as determined using the spectral breaks visible on days 426, 490, and 587. Mass estimates vary based on the density profile assumed.}
  \label{tab:Mass}
\begin{tabular}{@{}ccccccc@{}}
  \hline
  
 {\bf Epoch}  &{\bf Day}&	$\v_{i}$ (km s$^{-1}$)& $\zeta= \v_i/\v_{\rm obs}$	&M ($M_{\odot}$), $n \propto 1/r^1$ &M ($M_{\odot}$), $n \propto 1/r^2$ & M ($M_{\odot}$),  $n \propto 1/r^3$ 	 \\ \hline

16	& 	426 & 360 $\pm$ 20		&	0.24 $\pm$ 0.01 	&  2.7	$\pm$ 0.3 $\times 10^{-4}$  &	1.6	$\pm$ 0.2 $\times 10^{-4}$	&	0.9	$\pm$ 0.2 $\times 10^{-4}$		\\
17	& 	490 & 210 $\pm$ 10		&	0.14 $\pm$ 0.01	& 	4.0	$\pm$ 0.4 $\times 10^{-4}$  &	1.6	$\pm$ 0.2 $\times 10^{-4}$	&	0.6	$\pm$ 0.1 $\times 10^{-4}$		 \\
18	&  	587	& 190 $\pm$ 20	&	0.13 $\pm$ 0.01	& 4.6	$\pm$ 1.6 $\times 10^{-4}$ &	1.7	$\pm$ 0.7 $\times 10^{-4}$	&	0.7	$\pm$ 0.3 $\times 10^{-4}$	 \\ \hline
\end{tabular}

\end{minipage}
\end{table*}

\indent

We obtained a first estimate of the distance to V1723~Aql by using our maximum extension of the remnant on day 832, when our image had the highest spatial resolution (smallest clean beam). Since by day 832, some of the outermost material had become optically thin and therefore the remnant might not have a sharp outer edge, we took the maximum size scale to be the extent along the major axis of the gaussian fit, down to a flux density of twice the RMS. When we compare the size scale of $251 \pm 7$~mas with the expectation for ejecta expanding with a maximum velocity of $\v_{\rm obs}=1500\;{\rm km}\,{\rm s}^{-1}$  \citep{Yamanaka10}, we obtain a distance of $5.7 \pm 0.4$ kpc, or $\sim$6~kpc. The assumption that the maximum expansion speed in the plane of the sky is equal to the maximum radial velocity is valid if the remnant is approximately spherical or, for a bipolar flow, has wide opening angles (as is suggested by the $1/r^2$ density profile, see below).

Other methods of distance determination using optical data are either not possible or extremely unreliable. There are two means of nova distance determination using optical data. The most common utilized method is the Maximum Magnitude Rate of Decline (MMRD) method \citep{DellaValle95}; the other method involves utilizing high-resolution spectra to identify absorption lines from intervening interstellar clouds of known distance \citep[e.g.][]{Chomiuk12}. No high resolution spectra of V1723~Aql were available, making it impossible for us to use the latter method. The MMRD technique has been the subject of serious critiques in recent publications \citep[e.g.][]{Kasliwal11, Cao12}. Further, the episode of dust formation observed in V1723~Aql confounds the determination of $t_2$, rendering the already dubious MMRD method useless for distance determination.

Using d$\sim$6~kpc, we find that the brightness temperature after day 200 was consistent with the late-time radio emission emanating from a thermal remnant with $T \sim 10^4$~K, where $T$ is the electron temperature. {Brightness temperature is}
\be
T_b(\nu, t) \sim \frac{S_\nu (t)c^2 D^2}{2 \pi k_b \nu^2 (\v_{\rm obs}t)^2},
\ee
\noindent
{where} $c$ is the speed of light, $k_b$ is Boltzmann's constant, $t$ is the time since eruption, and $ \v_{\rm obs}$ is the observed maximum radial velocity \citep[e.g.][]{Bode08}. {Using our values from day 200, for example, we find that} $T_{\rm b} = 1.5 \pm 0.4 \times 10^4$~K {at 5.2~GHz}, {which is consistent with} $\sim$10$^4${~K.} Ionized ejecta from novae are typically found to have electron temperatures of around $10^4$~K, which is also approximately the equilibrium temperature for photoionized plasma that is cooled by forbidden line emission \citep{Seaquist77, Taylor87, Bode08}. This $\sim$10$^4$~K ejecta temperature can be sustained by photoionization heating by the WD for up to a year after the nova eruption \citep{Cunningham15}. A brightness temperature around this value therefore provides support for thermal bremsstrahlung as the dominant emission mechanism at late times in V1723~Aql.

Moreover, the detailed evolution of the radio spectrum between days 198 and 350 (approximately 6 to 12 months after the start of the outburst) places constraints on the density profile in the outermost regions of the expanding ejecta. We consider the point at which the free-free optical depth at a given frequency ($\tau_\nu$) is unity, where the emitting material transitions from optically thick to optically thin. The high spectral index of $\alpha_{\rm low} \approx 1.5$ on day 198 indicates that at that time, the ejecta were dense enough to be optically thick for frequencies of up to at least 36.5~GHz. Therefore, these $\tau_\nu = 1$ surfaces were very close to the outer boundary of the ejecta for all frequencies at which we observed on day 198. The appearance of the first spectral break on day 325, however, reveals that by that time, the density of the ejecta had dropped enough that at least part of the ejecta were beginning to become optically thin for frequencies above $\nu_{\rm lt}$. Thus, the radii of the $\tau_\nu = 1$ surfaces at frequencies above $\nu_{\rm lt}$ were no longer increasing as fast as that of the ejecta. The spectral index above the first spectral break ($\alpha_{\rm  trans}$) is directly related to the density profile in these transitioning ejecta. For an isothermal shell with a density profile of the form $n(r) \propto 1/r^\beta$ (where $n$ is density and $r$ is distance from the central binary), the spectral index is given by $\alpha_{\rm trans} \sim \frac{4\beta-6.2}{2\beta-1}$ for $\beta \geq 1.5$ \citep{Wright75}. For a measured $\alpha_{\rm trans}$, we thus have $\beta=\frac{31-5\alpha_{\rm trans}}{20-10\alpha_{\rm trans}}$ (see Table~\ref{tab:SpecIndex}). Our measurement of $\alpha_{\rm trans} \approx 0.6$ on days 325 and 350, when only the outermost portions of the ejecta were beginning to become optically thin, suggests that the density profile in the outermost part of the ejecta was proportional to $1/r^2$.  

The appearance of a second break in the spectrum on day 426, roughly 14 months after the start of the eruption, shows that the ejecta had become optically thin at frequencies above $\nu_{th}$. Optically thin thermal emission has a spectral index of $\alpha = -0.1$ \citep[e.g.][]{Bode08},as we approximately found above $\nu_{th}$. Between the two breaks at $\nu_{lt}$ and $\nu_{th}$, the spectral index of the transitioning region $\alpha_{\rm trans}$ provides information about the density profile deeper within the ejecta. With a value of approximately 0.3 on days 490 and 587, when these two breaks are clear, $\alpha_{\rm trans}$ suggests an average density profile deep within the remnant of roughly $1/r^{1.7}$ -- somewhat flatter than in the outermost regions of the remnant.  

The observed evolution of $\alpha_{\rm trans}$ is unlikely to be due to a temperature gradient. Although a temperature gradient can in principle affect the spectral index, the effect drops out for a $1/r^2$ density profile \citep{Cassinelli77, Barlow79}. Moreover, a radial decrease in the ionized gas fraction would also only serve to increase the steepness of the spectrum during the transition period \citep{Wright75}, in contrast to what we observed. Therefore, the observed flattening of the $\alpha_{\rm trans}$ in the interior of the ejecta is unlikely to have been due to a temperature or ionization gradient.  

For a more promising avenue to explain the declining $\alpha_{\rm trans}$, \citet{Wright75} note that non-spherical geometries can cause the total number density of the gas to fall off less rapidly than $n\propto r^{-2}$ for a constant, finite duration flow. For example, a disk geometry would fall off as $n \propto r^{-1}$ and in a cylindrical geometry $n$ would be independent of $r$ \citep{Wright75}. Therefore, that for V1723, Aql $\alpha_{\rm trans}$ appears to flatten slightly with time may indicate that the inner remnant is more confined towards the equatorial plane than the outer remnant  -- as in the case of the biconical flow collimated by a dense, slow torus in V959~Mon \citep{Chomiuk14}. If this is the case, a bi-polar outflow could result in an over-estimate of mass by up to a factor of two \citep{Ribeiro14}. Nevertheless, since the density profile of 1/$r^2$ in the outer regions of the ejecta suggest that any bipolar flow must have a wide opening angle, we can nevertheless use a spherical model to obtain reasonable estimates of the ejecta parameters.

The flux density {near} $\nu_{\rm th}$ furnishes a rough estimate of the inner
radius of the ejecta on the dates when this break frequency was
observable. {The ejecta are optically thin, with optical depth} $\tau_\nu <1$, {at frequencies above }$\nu_{\rm th}$. {They are optically thick, with}
$\tau_\nu > 1$ {at frequencies below} $\nu_{\rm th}$.
{Integrating the absorption coefficient for radio emission at}
$\nu_{\rm th}$ {along the line of sight through the ejected envelope therefore
gives exactly} $\tau_\nu = 1$, {suggesting that the photosphere lies near the inner edge of the ejected shell for $\nu_{\rm th}$. At the frequency} $\nu_{\rm th}$, {the rough angular size of the photosphere is}

\be
\frac{\theta}{\rm arcsec} = \Bigg(\frac{S_\nu}{\rm mJy}\Bigg)^{1/2} \Bigg(\frac{T_b}{1200\;{\rm K}}\Bigg)^{-1/2} \Bigg(\frac{\nu_{\rm th}}{\rm GHz}\Bigg)^{-1}
\ee
 \citep{Rybicki79}, where $\theta$ is the full angular diameter of the photosphere. Using the break frequencies $\nu_{\rm th}$ and flux densities at these frequencies from days 426, 490, and 587 (see Fig.~\ref{fig:LateSpec}), and taking $T_b \approx 10^4$~K, we find that the inner edge of the ejecta had an approximate radius of between 10 and 15~mas during these observations, which corresponds to an expansion speed $v_i$ {of the slowest, innermost ejecta} of between 200
and 400~km~s$^{-1} (d/6~{\rm kpc})(T_b/10^4~{\rm K}),^{-1/2}$, assuming all the ejecta
were homologous and ejected at $t_0$. Although these estimates do not provide any evidence that the inner radius increased between days 426 and 587, the estimates are uncertain enough that they do not place meaningful constraints on the actual evolution of the inner radius. The estimates do, however, indicate that the radio-emitting ejecta were geometrically thick (given that the outermost material moved away from the central binary at approximately 1500~km~s$^{-1}$), though not uniquely or unusually so \citep[e.g.][]{Seaquist77}.

The radio spectra during the second year of the eruption also enable us to make an estimate of the ejecta mass. We can obtain the emission measure of the ejecta at the $\tau_\nu = 1$ radio photosphere at $\nu_{\rm th}$ from the expression
\be
 \tau_\nu = 8.235 \times 10^{-2}\,
\Bigg{(}\frac{T}{\rm K}\Bigg{)}^{-1.35}
\Bigg{(}\frac{\nu}{\rm GHz}\Bigg{)}^{-2.1}\Bigg{(}\frac{EM_{rad}}{\rm cm^{-6}\;   pc}\Bigg{)}
  \ee
where $\tau_\nu$ is the free-free optical depth at frequency $\nu$ and $EM_{rad}=\int_{R_i}^{R_o} n^2(r) \; dr$ is the emission measure integrated from the inner radius $R_i$ to the outer radius $R_o$ \citep{Lang80}. With the emission measure, inner and outer radii, and density profile in hand for the observation with the clearest break frequency $\nu_{\rm th}$, we can estimate the mass of the ejecta. For the purpose of this mass estimate, we take a spherically symmetric shell and density proportional to $1/r^2$, as estimated from $\alpha_{\rm trans}$ in the outer regions of the ejecta. Using the inner radius on day 426 from above, which gives the velocity of the inner edge as v$_i =360$ km s$^{-1}$ and outer radius assuming a maximum expansion velocity of 1500~km~s$^{-1}$, we find an approximate ejecta mass of $1.6( \pm 0.2)\times 10^{-4}$ M$_\odot$, which is of the same order of magnitude as is typically found in nova ejecta mass estimates \citep{Bode08}. Repeating this process for other days on which multiple breaks were visible results in similar mass estimates -- on day 490 we find a mass of $1.6( \pm 0.2)\times 10^{-4}$ M$_\odot$, and on day 587 we find a mass of $1.7( \pm 0.7)\times 10^{-4}$ M$_\odot$ (see Table~\ref{tab:Mass}).

\subsubsection{The Hubble Flow model}

\begin{figure*}
\begin{minipage}{180 mm}
\centering
\caption{The late time observations of V1723~Aql between 2011 March and 2014 March, presented here for the first time, with eight previously published data points from days 2011 January 31 and 2011 March 5 used to constrain the high frequency light curves. Model indicates light curve of an expanding spherical thermal shell with a density profile of $1/r^2$, mass M=$1.7\times 10^{-4}$ $M_{\odot}$, ejecta temperature of T=$1.1 \times 10^4$ K, maximum velocity of $\v_{\rm obs}$=1500 km s,$^{-1}$ distance of d= 6.0 kpc, and ratio of minimum to maximum velocity $\zeta=\v_i/\v_{\rm obs} = 0.17$.}
\includegraphics[width=180mm]{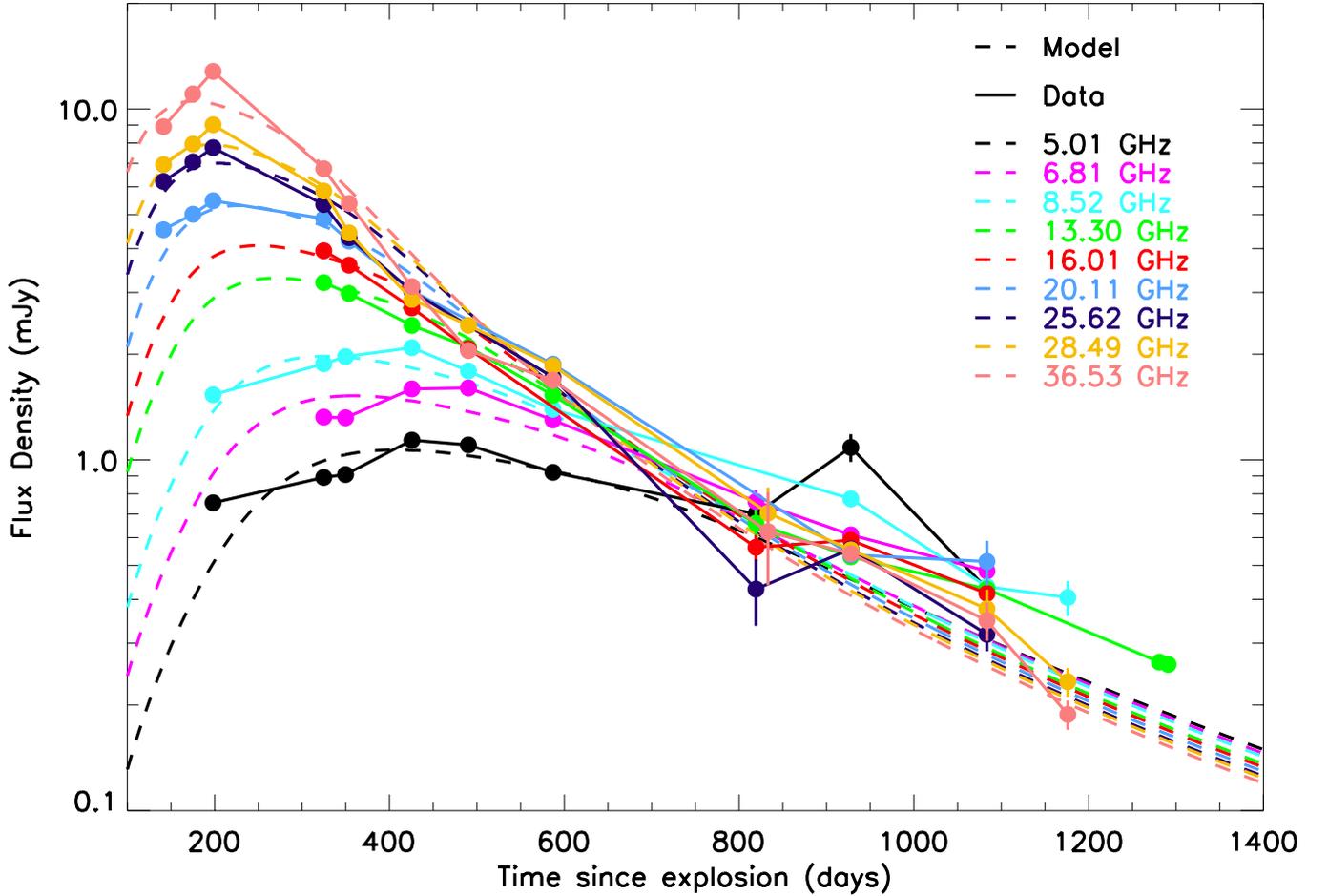}
\label{fig:NewModelFlux} 
\end{minipage}
\end{figure*}

\begin{figure*}
\begin{minipage}{180 mm}
\centering
\caption{{Variations} of the Hubble Flow model compared to the observed flux densities. Unless otherwise stated, the parameters for the model show distance of 6 kpc, maximum velocity of 1500 km s$^{-1}$, temperature of $1.1 \times 10^4$ K, mass of $1.7 \times 10^{-4} M_\odot$, and $\zeta=0.17$. The leftmost column shows variation in temperature, with the top panel showing a temperature of $3 \times 10^4$ K and the bottom showing a temperature of $5 \times 10^3$ K. The middle column shows variation in mass: the top panel shows a model with mass $2.5 \times 10^{-4}$ $M_\odot$, and the bottom panel shows $1.2 \times 10^{-4}$ $M_\odot$. The right column shows variation in shell thickness, varying the ratio of minimum to maximum velocity $\zeta$. The top panel shows a ratio of 0.4, and the bottom shows a ratio of 0.08.}
\includegraphics[width=180mm]{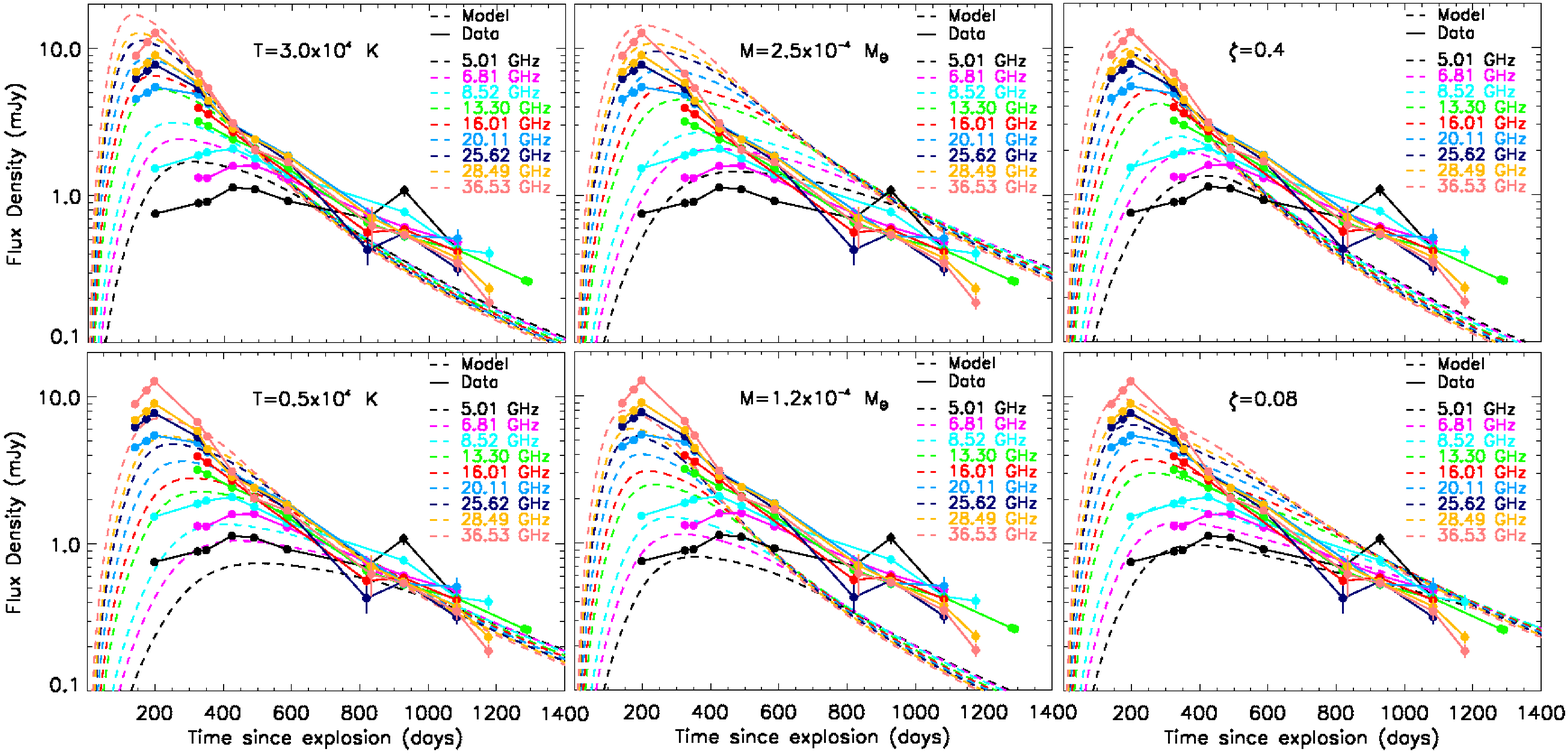}
\label{fig:ModelRange} 
\end{minipage}
\end{figure*}

\indent 

The gross properties of the radio-emitting ejecta at late times (after day 200) were consistent with expectations from the so-called ``Hubble-Flow'' model \citep{Seaquist77, Hjellming79}. In this model, a spherically symmetric shell is ejected from the WD instantaneously, but with a range of velocities, so that the velocity is proportional to the distance from the WD. A $1/r^2$ density profile, as we inferred from $\alpha_{\rm trans}$, corresponds to the mass being distributed uniformly over the velocities. Such a model has been applied to many historical novae, including V1974 Cyg \citep{Hjellming96}, V1500 Cyg \citep{Seaquist80}, QU Vul \citep{Taylor88}, and V723 Cas \citep{Heywood05}. Given the brightness temperature after day 200 of near $10^4$~K, and the evolution of the radio spectrum from initially rising with frequency (as expected for an optically thick shell) to falling slightly with frequency (as expected for an optically thin shell) via the appearance of two distinct spectral breaks, we conclude that it is reasonable to use a Hubble-flow model to parametrize the ejecta that produced the late-time radio emission from V1723~Aql.  

\indent

Although the Hubble-flow model does not give a formally acceptable fit to the late-time radio emission, it is roughly consistent with the most dominant features in the data. Fig.~\ref{fig:NewModelFlux} shows the Hubble-flow model light curves that provide the best fitting to the late-time radio data (where we have added eight data points published in \citet{Krauss11} from days 141 and 175 to our measurements to help constrain the turnover of the high-frequency light curves). Taking the nova to have a distance of 6~kpc and using the expansion velocity measured by \citet{Yamanaka10} on the day of discovery as the maximum velocity $\v_{\rm obs}=1500$ km s$^{-1}$, ejected at $t=0$, and allowing other parameters to vary, our best fitting model has an ejecta mass of $M = 1.7 \times 10^{-4}$ $M_\odot$, a temperature of $T = 1.1\times10^4$~K, and a ratio of minimum to maximum velocity $\zeta=$\v$_i/$\v$_{\rm obs} = 0.17$ (see Fig.~\ref{fig:NewModelFlux}). Given that the ejecta are clearly not spherically symmetric (see Fig.~\ref{fig:Resolved}), and that the evolution of the spectral index $\alpha_{\rm trans}$ from 0.6 to 0.3 suggests a flattening of the density profile over time, it is not too surprising that the Hubble-flow model does not provide a formally acceptable fit. Overall, however, the fitting of the late-time radio data with a Hubble-flow model indicates that if the ejecta were approximately spherical, then roughly $2\times 10^{-4}\, $ M$_\odot$ was ejected at a temperature of $10^4$~K.

Comparing model light curves for a range of ejecta masses, temperatures, and shell thicknesses provides a measure of the uncertainty on these parameters. Fig.~\ref{fig:ModelRange} suggests that the best fitting parameters are accurate to within at least a factor of 2. In terms of the effect that modifying these parameters has on the model light curves, changing the distance would have influenced the maximum flux density at each frequency, to which it is inversely related. Increasing the temperature of the shell increases the resulting flux density at all times, and causes the turnover from optically thick to thin to occur earlier. Increasing the maximum velocity also causes this transition to occur at earlier times. The ratio of the minimum and maximum velocity, $\zeta$, relates to the flux density and the decline of the light curve after peak -- a higher ratio corresponds to a geometrically thinner remnant, which both raises the maximum flux density and causes a swifter decline in flux density after this peak. Increasing the total ejecta mass in the model also causes an increase in flux density at all times, and causes a slower, later transition from optically thick to optically thin emission at all frequencies. 

While our observations are generally consistent with a $n \propto r^{-2}$ density profile, we can additionally demonstrate the sensitivity of the total ejecta mass to the distribution of mass within the nova shell. For example, if we again use the observed spectral breaks to estimate the inner and outer radii of the shell and calculate the emission measure, we can estimate the mass of the shell with different density profiles. For example, by using the spectral breaks on day 587, we estimate an ejecta mass of $M\sim 4.6 \times 10^{-4} M_{\odot}$ and $M\sim 0.7\times 10^{-4} M_{\odot}$ for densities profiles of $1/r$ and $1/r^3$, respectively (see Table~\ref{tab:Mass}). If the ejecta from V1723~Aql were clumpy rather than smooth, the total ejecta mass would be reduced. Clumping increases the flux density of an isothermal shell by up to a factor of $f^{-2/3}$, where $f$ is the volume filling factor \citep{Abbott81}. However, clumping would not influence the spectral index, nor the overall shape of the radio light curve, as uniformly distributed clumping would effect the free-free flux similarly at all frequencies \citep{Abbott81, Leitherer91, Nelson14}.

\subsection{Origin of the early time flare: particles accelerated in shocks within the ejecta} 
\subsubsection{Brightness temperature of the ejecta}

\begin{figure*}
\centering 
\begin{minipage}{165 mm}
\caption{Flux densities (top panel) and brightness temperatures
  (bottom panel) for V1723~Aql
  between 2010 September and 2014 March. The dashed lines indicate the
  expected thermal emission from an expanding spherical shell with a
  density profile proportional to $1/r^2$, mass M=$1.7\times 10^{-4}
  \; M_{\odot}$, ejecta temperature of T=$1.1 \times 10^4$~K, maximum
  velocity of $\v_{\rm obs}$=1500~km~s$^{-1}$, distance of d =
  6.0~kpc, and ratio of minimum to maximum velocity
  $\zeta=\v_i/\v_{\rm obs} = 0.17$.  The brightness temperature, which
  is an expression of surface brightness, is
  highly frequency dependent and peaks 
during the early-time flare at all frequencies.  The great difference
between the brightness temperatures during the early-time flare and
the late-time peak highlights the physically distinct
nature of the early and late radio brightenings.
}
 \includegraphics[width=165 mm]{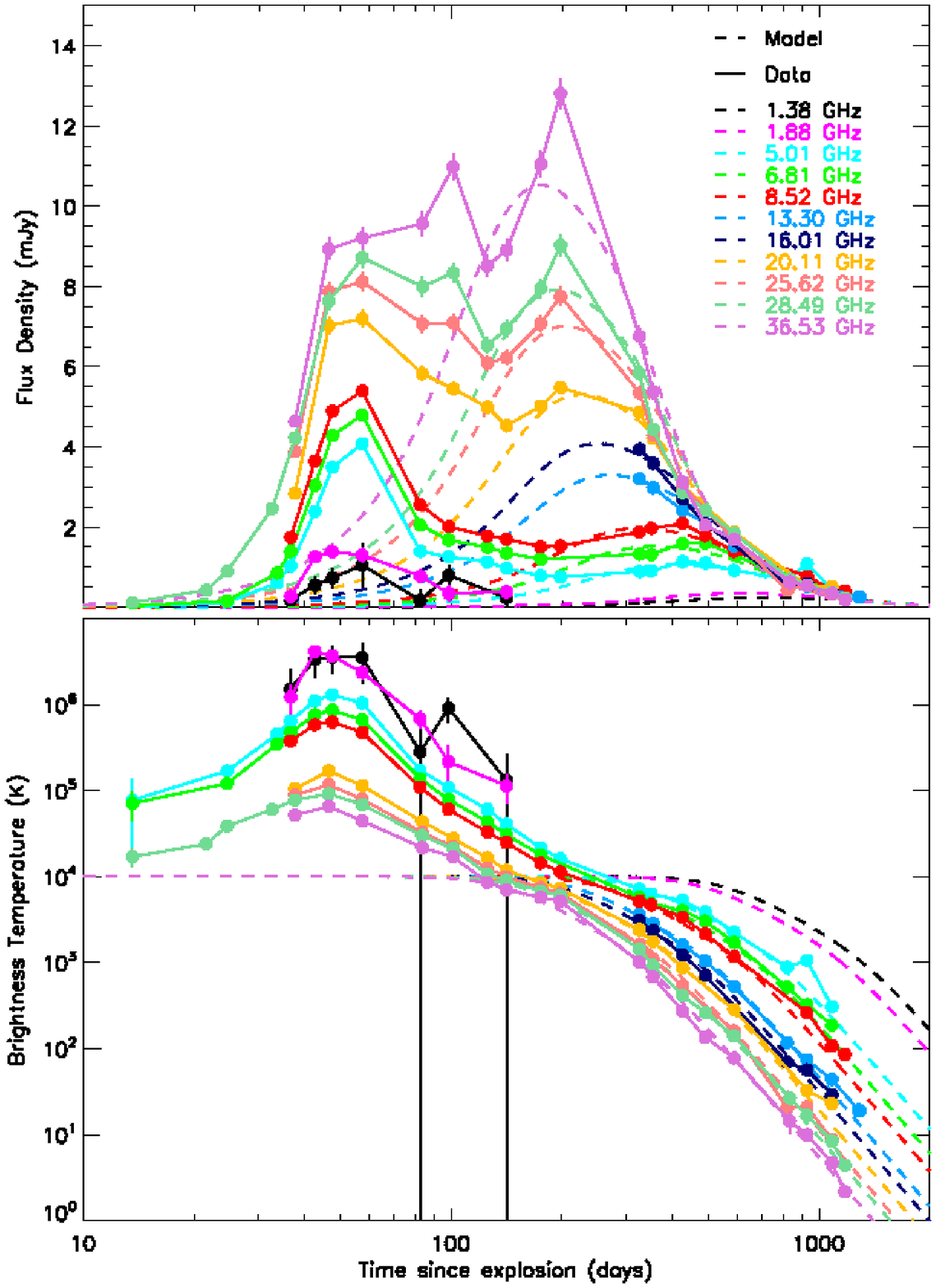}
\label{fig:FluxTemp} 
\end{minipage}
\end{figure*}

\indent

Whereas the late-time emission is consistent with originating in a
thermal remnant with $T \sim 10^4$~K and a strong resemblance to a
Hubble flow, the early-time emission is not (see
Fig.~\ref{fig:FluxTemp}). In fact, the $T \sim 10^4$~K ejecta were
only revealed once the early-time flare faded. The brightness
temperature at early times was strongly dependent on both frequency
and time, with a maximum $T_b$ of between $\sim 6\times 10^4$~K at
36.5~GHz and $\sim 4 \times 10^6$ K at 1.4~GHz. Brightness temperature
is an expression of surface brightness and represents a lower limit on
the physical temperature of any thermally-emitting material; for the
flare to have been thermal, it would therefore have had to have been
emitted by plasma with $T \geq 4 \times 10^6$~K. 

During the early-time flare, however, the X-ray flux was too low for
the radio flare to have been due to hot (X-ray emitting) plasma.
For the radio flare to have been entirely
thermal, it would have had to have been produced by plasma that was
optically thick at 1.9~GHz, had an electron temperature $T \approx 4
\times 10^6$~K, and had a size scale of $3 \times 10^{-4}{\text{ pc }}
\Big(\frac{\v_{\rm obs}}{1500\,{\rm\; km \; s}^{-1}}\Big)\Big(\frac{
  t}{43\,{\rm d}}\Big)$. The flare could in principal also have been
produced by free-free emission from hotter plasma that subtended a
smaller solid angle on the sky. \citet{Krauss11} detected X-ray
emission from plasma with $T > 2 \times 10^7$~K on day 41 after the
start of the eruption.   For
ejecta with  $T= 2 \times 10^7$~K to be optically thick, the free-free
optical depth would have to have been greater than one: 
\begin{equation}
 \tau_{ff} = 3.0 \times 10^{-12}\; \Bigg(\frac{T}{2 \times
   10^7\;{\rm K}}\Bigg)^{-1.35} \Bigg 
(\frac{\nu}{1.9\, {\rm GHz}}\Bigg)^{-2.1}
 EM_{\rm rad} > 1.
\end{equation}
\noindent
This requirement corresponds to
\begin{equation}
 \int n_e^2 dl > 3.3\times10^{11} \; \Bigg(\frac{T}{2\times 10^7 \; {\rm K}}\Bigg)
^{1.35} \Bigg(\frac{\nu}{1.9\;{\rm GHz}}\Bigg)^{2.1} \text{pc cm.}^{-6} 
\end{equation}

\noindent
A hot plasma with a temperature five times the maximum $T_b$ could in principal have produced the radio flare if it was optically thick at 1.9~GHz and subtended a solid angle that was 20\% that of the entire ejecta. A uniform-density spherical shell of that size, with shell thickness $\ell$, would therefore require an electron density of 

\begin{eqnarray}
n_e > 6 \times 10^7 & \!\!\!\!\times \;\; \Big(\frac{T}{2 \times 10^7 \; K}\Big)^{0.675} \Big(\frac{\nu}{1.9 \;{\rm GHz}}\Big)^{1.05}\Big( \frac{\ell}{2.5 \times 10^{14}\; {\rm cm}}\Big)^{-1/2} {\rm cm}^{-3}
\end{eqnarray}

\noindent
to be optically thick. Such hot plasma would produce an X-ray emission measure, $EM_{\rm x} \equiv \int n_e^2 dV$ (where $V$ is the emitting volume), of approximately $8 \times 10^{59}\; {\rm cm}^{-3}$ (independent of $\ell$), and (unless the shell was extremely thin) have a mass of a few times $10^{-6}$~M$_\odot$. On day 41, however, $Swift$/XRT detected X-rays with an $EM_{\rm x}$ of just $\sim 10^{56}\; {\rm cm}^{-3}$, indicating that there  was not enough $2\times 10^7$~K plasma to produce the observed radio flare. Even generating a model of the expected X-ray emission and assuming the highest reasonable absorbing column allowed by the X-ray spectral fit ($3 \times 10^{23}$~cm$^{-2}$), the observed X-ray flux was still an order of magnitude lower than would be required to explain the radio flare with thermal emission. The natural alternative to thermal emission is radio synchrotron emission from electrons accelerated in a shock.

The inconsistency between the radio flux density at 1.9~GHz and thermal radio emission expected from the $2\times10^7$~K X-ray emitting plasma is independent of the distance to V1723~Aql. If we were to allow the temperature of the hot plasma to drop below $2\times10^7$~K (despite the constraints from \citet{Krauss11}), then the evidence for synchrotron emission would in principle depend on the distance to V1723~Aql. However, for the distance to be small enough to explain the radio flare with thermal emission, the outflow velocities become inconsistent with constraints from optical spectroscopy. Therefore, the combination of radio, X-ray, and optical observations appear to support a synchrotron emission mechanism. 

\subsubsection{Failure of thermal shock emission model}

\begin{figure}
\begin{minipage}{\columnwidth}
\centering
\caption{Observed flux densities and model flux densities in radio for a thermal shock. Assuming a distance of 6 kpc, the model uses a fast wind with an initial velocity $\v_f = 2100 {\rm \; km \; s}^{-1}$, mass $M_w =  0.8 \times 10^{-4}$ $M_\odot$, and a characteristic duration of 12 days. The wind collides with a slower moving circumbinary shell with initial velocity $\v_s =  1300{\rm \; km \; s}^{-1}$ and mass $M_s=1 \times 10^{-4}$ $M_\odot$. This results in a post shock velocity of $\v_{\rm obs}=$1500~km~s$^{-1}$ and total nova ejecta mass as $=  1.8 \times 10^{-4} M_\odot$, consistent with our late-time thermal model.}
 \includegraphics[width=\columnwidth]{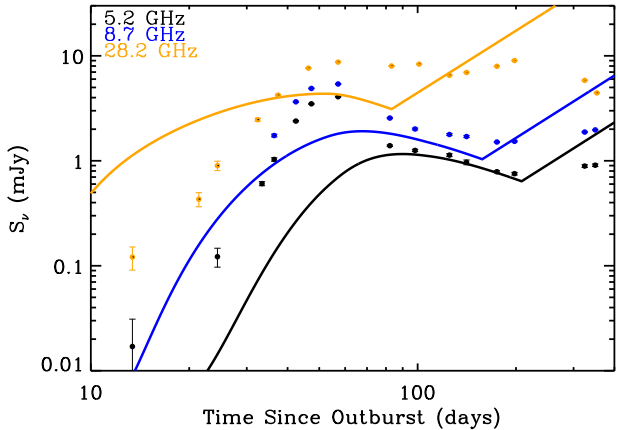}
\label{fig:ThermalShock} 
\end{minipage}
\end{figure}

\indent

To further assess whether the early-time radio flare in V1723~Aql was thermal or non-thermal (synchrotron) in origin, we consider a one-dimensional, spherically-symmetric model for internal shocks in novae, as presented in \citet[][hereafter M14]{Metzger14}. Consider a fast outflow of velocity $\v_{\rm f} \gtrsim 10^{3} {\rm \; km \; s}^{-1}$ that collides with a slower shell of velocity $\v_{\rm s} \lesssim \v_{\rm f}$. For simplicity, assume that the mass of the slow shell is comparable to or lower than that of the fast material and is emitted at time $t = 0$. 
The fast outflow is emitted a time $\Delta t$ after the slow shell, resulting in a collision radius 
\be
R_{\rm c} \approx \frac{\v_{\rm s}\v_{\rm f}\Delta t}{\v_{\rm f}-\v_{\rm s}} \sim 6\times 10^{13}\; \v_{\rm f,8}\frac{\v_{\rm s}}{\Delta \v}\left(\frac{\Delta t}{\rm week}\right){\rm \, cm\,}
\ee
and time
\be
t_{\rm c} \approx \frac{R_{\rm c}}{\v_{\rm s}}  \approx \frac{\v_{\rm f}}{\Delta \v}\Delta t
\ee
where $\Delta \v \equiv \v_{\rm f}-\v_{\rm s}$ and $\v_{\rm f,8} = \v_{\rm f}/10^{8}$~cm~s$^{-1}$.  

The predicted radio flare following this collision would be powered by the forward shock that emerges from the slow shell at approximately the time of the collision, i.e. $t_{\rm c} \sim 50$~days. If both shells are ejected over a timescale less than the duration of the peak optical activity, i.e.~$\Delta t = (\Delta \v/\v_{\rm f})t_{\rm c} \lesssim 2$ weeks, then the relative velocity between the shells must be relatively low, $\Delta \v \lesssim \v_{\rm f}(\Delta t/t_{\rm pk}) \approx 0.3 \v_{\rm f}$, corresponding to a collision radius $R_{\rm c} \sim \v_{\rm s}t_{\rm c} \lesssim \v_{\rm f}t_{\rm c} \sim 10^{15}$~cm for $\v_{\rm f} \sim 2\times 10^{3}{\rm \; km \; s}^{-1}$.  

The forward shock heats gas to a temperature (M14, eq.~31),
\be
T \approx 1.4\times 10^{7}\left(\frac{\Delta \v}{10^{3}\,{\rm km\,s^{-1}}}\right)^{2}{\,\rm K} \approx 1.3\times 10^{6}\left(\frac{\Delta \v}{0.3\v_{\rm f}}\right)^{2} \v_{\rm f,8}^{2}\,{\rm \; K}
\label{eq:T}
\ee
corresponding to an X-ray energy of $kT \sim 0.5$~keV for characteristic values $\Delta \v \sim 0.3 \v_{\rm f}$ and $\v_{\rm f} \sim 2\times 10^{3}{\rm \; km\; s}^{-1}$. {For purposes of an estimate, here we have assumed that the slow shell is of comparable or lower density than the fast outflow. }

{For} $n \approx 10^{7}$ cm$^{-3}$ and $kT \approx 0.5$ {keV the cooling time of the gas is approximately one week for solar abundances, and even less for} $X_Z \sim 0.1$ {as assumed below.  This suggests that the shock is likely to be radiative, in which case the resulting X-ray luminosity is given by}
\begin{eqnarray}
L_{\rm X} &\approx& \frac{9\pi}{32}n m_p R_{\rm c}^{2}(\Delta \v)^{3} \nonumber \\
&\approx& 4\times 10^{{34}}\,\left(\frac{n}{{10^{6}}\,{\rm cm}}\right)\left(\frac{R_{\rm c}}{10^{15}\,{\rm cm}}\right)^{2}\left(\frac{\Delta \v}{0.3\v_{\rm f}}\right)^{3} \v_{\rm f,8}^{3} {\rm\; erg\,s^{-1}}
\label{eq:Lx}
\end{eqnarray}
{where} $n$ is the (unshocked) density of the slow shell at the time of the collision. That the observed X-ray luminosity of $L_{\rm X} \sim 3 \times 10^{32}$~erg~s$^{-1}$ near the radio peak \citep{Krauss11} is {several} orders of magnitude smaller than the fiducial value in equation
(\ref{eq:Lx}) would require strong absorption of these soft X-rays by
neutral gas ahead of the shock. 
Indeed, the bound-free X-ray optical depth ahead of the shock (M14,
their eq.~50)   
\be
\tau_{\rm X} \approx 1.2\left(\frac{X_{\rm Z}}{0.1}\right)\left(\frac{n}{10^{7}\,{\rm cm^{-3}}}\right)\left(\frac{R_{\rm c}}{4\times 10^{14}\,{\rm cm}}\right)
\ee
exceeds unity for $n \gtrsim 10^{7}$ cm$^{-3}$, where $X_Z = 0.1$ is
the assumed mass fraction of  CNO elements. This density corresponding
to a minimum mass of the slow shell of $M_{\rm sh} \gtrsim 4\pi
(R_{\rm c}^{2}\Delta) n m_p \sim 10^{-4}\,M_{\odot}$ for $R_{\rm c}
\sim 10^{15}$ cm.  {For values of} $n$ {below which} $\tau_x < 1$, $L_X$ {in equation} (\ref{eq:Lx}) {still exceeds the observed X-ray luminosity by at least an order of magnitude; this may suggest that} $v_{f,8}\lesssim 1$, {or that a fraction of the shock power instead emerges at UV/soft X-ray energies below the {\it Swift} bandpass.}

The maximum radio brightness temperature of radiative shocks at frequency $\nu$ is given by (M14, eq.~65),
\begin{eqnarray}
 \qquad T_{\rm b,\nu}^{\rm max} \approx
\end{eqnarray}
\begin{eqnarray*}
 \left\{
\begin{array}{ll}
 3\times 10^{5} \; {\rm ln\left[\frac{T}{T_{0,\nu}}\right]}{\,\rm K} \;\; \underset{T \sim 10^{6}-10^{7}\rm \; K}   \sim \; 10^{6}{\rm \; K}
& n \gtrsim n_{\rm th} \\
T_{0,\nu} = 10^{5}\; \nu_{10}^{-2}\, n_7\left(\frac{\Delta \v}{\v_{\rm f}}\right)\; \v_{\rm f,8}{\;\rm K} \; \approx 4\times 10^{5} \; \nu_{10}^{-2}\left(\frac{n}{n_{\rm th}}\right){\rm \; K}     
 & n \lesssim n_{\rm th}, 
\end{array}
\right.
\label{eq:Tobsrad}
\end{eqnarray*}
where $\nu_{10} = \nu/10$ GHz, $n_7 = n/10^{7}$ cm,$^{-3}$ and 
\begin{equation}
n_{\rm th} \approx 4\times 10^{7}\left(\frac{\Delta \v}{0.3\v_{\rm f}}\right)^{-1} \v_{\rm f,8}^{-1} \; \nu_{10}^{2}\,{\rm cm^{-3}}
\label{eq:ff}
\end{equation}
defines the pre-shock density above which the layer of X-ray ionized gas ahead of the shock is optically thick to free-free absorption, where a temperature of $2\times 10^{4}$~K is assumed for the pre-shock gas.  {This value, appropriate to the photo-ionized layer slightly ahead of the shock} (\citealt{Metzger14}), {may be slightly higher than the values expected for the bulk of the ejecta later, which our Hubble flow modeling finds is closer to} $1\times 10^{4}$ K.  

The observed peak brightness temperature of V1723~Aql (see Fig.~\ref{fig:FluxTemp}) exceeds the range given by equation (\ref{eq:Tobsrad}) by more than order of magnitude, strongly suggesting a non-thermal origin for most of the observed radio emission. Furthermore, equation~(\ref{eq:Tobsrad}) was derived under the assumption that free-free emission is the sole source of cooling behind the shock. {If the effects of line cooling are included, Equation}~(\ref{eq:ff}) {greatly overestimates the brightness temperature}, further enhancing the tension with thermal emission models.  

It is difficult to produce a thermal model that fits the early time
radio data and is also consistent with the X-ray and late radio
constraint. In Fig.~\ref{fig:ThermalShock}, we applied the M14 model
to the early light curve, modeling the shock as the collision of a
fast wind with $\v_f = 2100{\rm \; km \; s}^{-1}$  and mass $ 0.8
\times 10^{-4}$ $M_\odot$ and a shell with $\v_s = 1300{\rm \; km \;
  s}^{-1}$ with mass $M_{sh} = 1 \times 10^{-4}$ $M_\odot$. We assumed
that the shells were released around t = 0, with a characteristic
duration of the wind of 12 days, motivated by the observed optical
decline rate. These parameters result in a post-shock velocity of
$\v_{\rm obs}=1500 {\rm\; km\; s}^{-1}$ and a total ejecta mass of $M
\sim 1.8 \times 10^{-4} M_\odot$, consistent with our late-time
thermal model. Taking $n \propto 1/r^2$ and distance of 6~kpc, we were
unable to reproduce the high radio flux densities over the course of
the shocks.  We conclude that synchrotron emission must therefore have
at least partially powered the early-time flare.

\subsubsection{Synchrotron emission from particles accelerated in shocks} 
\indent

The failure of thermal emission to account for the early-time radio flare suggests that the flare was dominated by synchrotron emission from particles that were accelerated in shocks. For one thing, shocks are common in classical novae. Evidence for shocks is provided by X-ray emission from plasma with temperature above $10^7$~K \citep[e.g.][]{Mukai08}. Such temperatures arise naturally behind shock with speeds on the order of 1000~km~s$^{-1}$ (see Equation~\ref{eq:T}). Furthermore, the recent identification of classical novae as a new class of $\gamma$-ray sources shows that not only are shocks ubiquitous in novae, but that relativistic particles are as well \citep{Ackermann14, Metzger15}. A natural acceleration mechanism for these particles is Fermi acceleration in the shocks. In fact, radio observations have illustrated the existence of radio synchrotron emission in a normal nova -- in {\it Fermi}-detected nova V959~Mon, \citet{Chomiuk14} identified the location of $\gamma$-ray producing relativistic electrons as the region where a fast outflow collided with, and was shaped by, a dense equatorial torus. Although V959~Mon did not produce a strong early-time radio flare, its early-time radio emission was consistent with being synchrotron \citep{Chomiuk14}. Whereas existing observations do not  constrain the inclination of V1723~Aql, V959~Mon is viewed very nearly edge on \citep{Chomiuk14} -- it is thus possible that some early non-thermal emission may have been obscured by the optically thick equatorial torus. Moreover, while the early-time flare in V1723~Aql suggests stronger non-thermal radio emission than that seen from V959~Mon, VLA observations of $\gamma$-ray producing nova V1324~Sco (Nova Sco 2012) revealed a double peaked radio light-curve similar to that of V1723~Aql (Finzell~et~al., in preparation).  

Moreover, synchrotron emission is physically plausible for V1723~Aql. We expect the magnetic field behind the shock to be amplified to a characteristic value \citep[e.g.][]{Caprioli14} 
\begin{eqnarray}
\label{eq:Bfield}
B & = & (6\pi \epsilon_B m_p n\Delta \v^{2})^{1/2} \nonumber \\
 &\approx & 0.05\epsilon_{B,-2}^{1/2}\left(\frac{n}{10^{7}\,\rm
  cm^{-3}}\right)^{1/2}\left(\frac{\Delta \v}{0.3\v_{\rm
    f}}\right)\v_{\rm f,8}{\; \rm G} 
\end{eqnarray}
where $\epsilon_{B} = 10^{-2}\epsilon_{B,-2}$ is the fraction of the post-shock thermal energy placed into turbulent magnetic fields. The characteristic synchrotron frequency from an electron of Lorentz factor $\gamma_e$ is given by 
\begin{eqnarray}
\nu_{\rm m} & \sim & \frac{eB \gamma_e^{2}}{2\pi m_e c} \nonumber \\
& \sim &1.4\,\; \epsilon_{B,-2}^{1/2}\left(\frac{n}{10^{7}\,\rm
  cm^{-3}}\right)^{1/2}\left(\frac{\Delta \v}{0.3\v_{\rm
    f}}\right)\v_{\rm f,8}\left(\frac{\gamma_e}{100}\right)^{2} {\rm \; GHz.}
\end{eqnarray}
For typical densities in the radio-producing nova shocks, $n \sim 10^{7}{\rm \; cm,}^{-3}$ the production of radio synchrotron thus requires relativistic electrons with $\gamma_e$, corresponding to energies of a few hundred MeV. Such electrons may be directly accelerated at the shock \citep[leptonic scenarios; see, e.g.,][]{Park15,Kato15}, or they may be created as  electron/positron pairs produced by the decay of charged pions (hadronic scenarios). The Larmor radii of the radio emitting particles ($\gamma_e$ of a few hundred) is extremely small compared to the size of the shock, so particles can easily remain within the acceleration region. {As a comparison, supernovae remnants, which typically have post-shock densities on the order of} ${0.1-10\rm \; cm,}^{-3}$ {and velocities of a few thousand km~s}$^{-1}$ \citep[e.g.,][]{Williams14, Metzger15b}, {using equation}~\ref{eq:Bfield} {would have magnetic field strengths scaling roughly as} 50~$\mu$G, {not atypical for post-shock remnants} \citep[as in, e.g.,][]{Ressler14}. {More detailed modeling of the non-thermal radio emission in novae (Vlasov~et~al., in preparation) find acceptable fits to the peak brightness temperature for assumed microphysical shock parameters} ($\epsilon_e \sim 10^{-3}-10^{-2}$,$\epsilon_B\sim 10^{-2}$) {as inferred from modeling supernova remnants} \citep[e.g., ][]{Morlino12}.

The properties of the radio flare are also consistent with a synchrotron contribution. The expected time scale for the decay of any synchrotron contribution is set by the size scale of the remnant divided by the expansion speed \citep{Taylor89}, which for V1723~Aql is around 40 days, as observed. \citet{Krauss11} did not detect linear polarization at 5.5~GHz on day 48, with a 3$\sigma$ upper limit of 39 $\mu$Jy (or 1\%). Similarly, on day 57, we detected no linear or circular polarization averaging across 1~GHz of bandwidth centred at 28.2~GHz, with 3$\sigma$ upper limits of 0.42~mJy (9\%) for circular polarization and 0.63~mJy (6\%) for linear polarization. Given that the magnetic field geometry was likely to have been complex, the lack of strong polarization is not a concern for the synchrotron picture. That the radio spectrum rose with frequency during the radio flare indicates that the synchrotron emission was either self-absorbed or suffered from free-free absorption. Since synchrotron self absorption would require the synchrotron-emitting region to have been extremely compact (i.e., have had a solid angle multiple orders of magnitude smaller than that of the remnant), whereas a reasonable amount of $10^4$~K ionized gas could have led to observable absorption, we find free-free absorption more likely.

\subsection{Nova Shocks in the context of bipolar flows and $\gamma$-rays}
\indent 

The perpendicular structures evident in Fig.~\ref{fig:Resolved} are reminiscent of the perpendicular flows that \citet{Chomiuk14} found to be responsible for the generation of shocks and gamma-rays in nova V959~Mon. In V959~Mon, radio observations showed that strong internal shocks produced synchrotron emission coincident with the $\gamma$-ray detection \citep{Chomiuk14}. Radio images of the ejecta showed rapidly expanding bipolar flows, which faded over the course of sixteen months to reveal an inner structure that collimated the outer flow  \citep[see][Fig. 2d]{Chomiuk14}. Our image from day 1281 (see Fig.~\ref{fig:Resolved}, panel b) reveals a similar inner structure that may be a shell with an equatorial density enhancement. The evolution of the radio spectrum of V1723~Aql may also indicate that the inner portions of its ejecta had a flatter density profile than the outer regions, as one might expect if an inner structure was more confined to the orbital plane (as in V959~Mon). Furthermore, the resolved image on day 832 showed two distinct components aligned roughly perpendicular to the axis of greatest elongation, which could be indicative of a limb-brightened collimating torus. 

By associating V1723~Aql's multiple, distinct radio structures with
different velocities, we identified flows that likely collided and led
to shocks and particle acceleration.
While optical spectra revealed the presence of a fast flow with maximum radial velocity $\v_{\rm obs}=1500{\rm \; km \; s}^{-1}$ \citep{Yamanaka10}, there is optical spectral evidence for slower moving ejecta as well. By using the break frequencies of the radio spectrum as the nova shell transitioned from optically thick to optically thin, we estimated that the slowest ejecta had an expansion speed of $\v_{i}$ between 200 and 400~km~s$^{-1} (d/6~{\rm kpc})(T_b/10^4~{\rm K})^{-1/2}$. The two components that were visible on day 832 had a separation of 70 mas on the plane of the sky, which, using a distance of 6~kpc, implies that each component had a velocity of roughly 440~km~s$^{-1}$ if they were ejected at $t=0$ -- consistent with our estimate for the innermost material. Additionally, \citet{Nagashima13} modeled optical emission lines on days 7 and 11 with 
two overlapping gaussian components with central velocities of -190~km~s$^{-1}$ and 360~km~s$^{-1}$ relative to line of sight, also  supporting a non-spherical distribution of ejecta with slow flows on the order of a few hundred km~s$^{-1}$. Finally, the expected temperature behind a shock between two flows with velocities of 440~km~s$^{-1}$ and 1500~km~s$^{-1}$ {is   }
$1.4 \times 10^7 \; \Big(\frac{\v_s}{1410 {\rm \; km\; s}^{-1}}\Big)^2  \; \rm{K}$ {(from equation} \ref{eq:T}), which is consistent with observational constraints on the temperature of the X-ray emitting plasma \citep{Krauss11}. That we measure a density profile close to $1/r^2$ from the transitioning ejecta indicates that the bipolar outflow likely had a wide opening angle. A model of V1723~Aql as a wide, biconical flow with collimating torus expanding at a few hundred km~s$^{-1}$ is consistent with imaging, optical and radio spectroscopy, the detection of X-ray emission, and the evolution of the radio light curve. The presence of more complex substructures or additional shocks could account for other variations in the radio light curve.

Fast biconical flows and  slow equatorial tori may be more common than
previously thought. For example, a change in the PA of maximum
elongation like that in V959~Mon and V1723~Aql was also evident in
Nova QU Vul, which had a double peaked light curve much like that of
V1723~Aql \citep{Taylor87, Taylor88}. Early imaging of QU Vul showed
bipolar structure, later evolving into a more spherical shape --
suggesting that as the outer bipolar emission became optically thin,
it revealed the central, more ring-like emission region
\citep{Taylor87}, as we saw in V1723~Aql.

That V1723~Aql displayed evidence for both shocks and dust production \citep{Nagashima13} around the time of the radio flare raises the intriguing possibility that dust production in novae is related to shocks. Both X-rays and radio synchrotron emission indicate that shocks were energetically important in V1723~Aql roughly 40$\pm$20 days into the eruption. \citet{Nagashima13} argued based on optical observations that dust was produced around day~20. The finding by \citet{Metzger14} that internal shocks generate a cool, dense post-shock region suggests that the appearance of shocks and dust at approximately the same time in V1723~Aql may be more than a coincidence.

\section{Conclusions}
\indent

Radio and optical imaging of novae has shown evidence for the
formation of jets, eg. RS Oph \citep{Sokoloski08}; rings, eg. DQ Her
\citep{Slavin95, Sokoloski13}; clumping, eg. T Pyx \citep{Shara97} and
V458 Vul \citep{Roy12}; collimating torus, eg. V959~Mon
\citep{Chomiuk14}; and other asymmetries. For V1723~Aql, the contrast
between the early- and the late-time radio emission reveals a system
of colliding flows in the ejecta. This finding has broader implications for the production of $\gamma$-rays and the shaping of the ejecta from novae. 

\begin{enumerate}

\item Late-time radio emission from V1723~Aql was roughly consistent
  with a freely expanding thermal plasma with a well-defined inner
  edge, a temperature of $10^4$ K, an ejecta mass of approximately $2
  \times 10^{-4} \; M_\odot$, a ratio of minimum to maximum velocity
  $\zeta=$\v$_i/$\v$_{\rm obs} \sim 0.2$, a maximum velocity of
  1500~km~s$^{-1}$, and a distance of 6~kpc.   

\item The early-time radio flare was most likely due to a combination
  of thermal and non-thermal synchrotron emission emanating from the
  shocks that produced the observed X-ray emission. Thermal shocks
  alone fail to reproduce the high brightness temperature and weak
  X-ray emission.  

  \item  The shocks that generated the early-time radio flare did not cause either the early optical light curve or the late-time radio evolution to significantly diverge from the expected behaviour of a standard fast nova. Relatively few novae have such comprehensive early-time radio observations, and thermal X-ray emission from shocks in novae is known to be common \citep{Mukai08} -- therefore, detectable radio emission from early shocks could be more ubiquitous than previously realized. 
  
\item It has been established that $\gamma$-ray production in the first few weeks of nova eruptions may be common \citep{Ackermann14, Metzger15}. $\gamma$-rays are presumably the result of particle acceleration in shocks, early radio flares -- as in V1723~Aql -- may provide an important diagnostic of the shocks and colliding flows that produce $\gamma$-rays. 
  
\item  Although the morpho-kinematic structure of the ejecta from
  V1723~Aql is not tightly constrained, its behaviour is consistent
  with that of a biconical flow loosely collimated by a {slow 
  equatorial} torus, as in V959~Mon \citep{Chomiuk14}. This hypothesis
  is supported by the existence of the perpendicular structures in the
  resolved radio images (see Fig.~\ref{fig:Resolved}). Moreover, the
  shock speed derived from the difference between the fast radio flow
  (1500~km~s$^{-1}$) and the slow radio flow (knots moving at $\sim$400~km~s$^{-1}$) is consistent with that inferred from the temperature of the post-shock X-ray emitting plasma. The flattening of the density profile seen from the variation in $\alpha_{\rm trans}$ could be a consequence of the material being slightly more concentrated towards the equatorial plane in the inner region of the remnant (see Fig.~\ref{fig:Cartoon}).
  
\end{enumerate}

 Radio observations are a powerful diagnostic for the colliding flows and shocks that are crucial for $\gamma$-ray production and shaping of the ejecta from novae. The radio emission from V1723~Aql shows that the collision of fast and slow flows resulted in a shock which, although leaving the late-time behaviour of the radio light curve mostly unaffected, produced both thermal and non-thermal emission. With the recent detections of classical novae in $\gamma$-rays, the question of relativistic particles in novae becomes even more important. \citet{Chomiuk14} found the morphology of the shock-producing structures in V959~Mon; our work here extends those findings to other classical novae.

\section*{Acknowledgments}
\indent

We thank NRAO for its generous allocation of time which made this work possible. The National Radio Astronomy Observatory is a facility of the National Science Foundation operated under cooperative agreement by Associated Universities, Inc. J. Weston and J.L. Sokoloski acknowledge support from NSF award AST-1211778. J. Weston was supported in part by a Student Observing Support award from NRAO (NRAO 343777). B.D. Metzger gratefully acknowledges support from NASA Fermi grant NNX14AQ68G, NSF grant AST-1410950, and the Alfred P. Sloan Foundation. L. Chomiuk, J. Linford, and T. Finzell are supported by NASA Fermi Guest Investigator grant NNH13ZDA001N-FERMI. T. Nelson was supported in part by NASA award NNX13A091G. We thank Benson Way for assistance with Fig~\ref{fig:Cartoon}.


\begin{thebibliography}{99}
\bibitem[\protect\citeauthoryear{Abbott, Bieging \& Churchwell}{Abbott et al.}{1981}]{Abbott81} Abbott D. C., Bieging J. H., Churchwell E., 1981, ApJ, 250, 645

\bibitem[\protect\citeauthoryear{Ackermann et al.}{2014}]{Ackermann14} Ackermann M. et al., 2014, Science, 345, 554

\bibitem[\protect\citeauthoryear{Balam, Hsiao \& Graham}{Balam et al.} {2010}]{Balam10} Balam D., Hsiao E. Y., Graham M., 2010, IAU Circ, 9168, 1 

\bibitem[\protect\citeauthoryear{Barlow}{1979}]{Barlow79} Barlow M. J., 1979, in IAU Symposium, Vol. 83, Mass Loss and Evolution of O-Type Stars, ed. P. S. Conti \& C. W. H. De Loore, 119-129

\bibitem[\protect\citeauthoryear{Bode \& Evans}{2008}]{Bode08} Bode M. F. , Evans A., 2008, Classical Novae (Cambridge University Press)

\bibitem[\protect\citeauthoryear{Cao et al.}{2012}]{Cao12} Cao Y. et al.,  2012, ApJ, 752, 133

\bibitem[\protect\citeauthoryear{Caprioli \& Spitkovsky}{2014}]{Caprioli14} Caprioli D., Spitkovsky A., 2014, ApJ, 794, 46

\bibitem[\protect\citeauthoryear{Cassinelli \& Hartmann}{1977}]{Cassinelli77} Cassinelli J. P., Hartmann L., 1977, ApJ, 212, 488

\bibitem[\protect\citeauthoryear{Chomiuk et al.}{2012}]{Chomiuk12} Chomiuk L. et al., 2012, ApJ, 761, 173

\bibitem[\protect\citeauthoryear{Chomiuk et al.}{2014}]{Chomiuk14} Chomiuk L. et al., 2014, Nature 514, 339

\bibitem[\protect\citeauthoryear{Cunningham, Wolf \& Bildsten}{2015}]{Cunningham15} Cunningham T., Wolf W. M., Bildsten L., 2015, ApJ, 803, 76

\bibitem[\protect\citeauthoryear{Della Valle \& Livio}{1995}]{DellaValle95} Della Valle M., Livio M.,1995, ApJ, 452, 704

\bibitem[\protect\citeauthoryear{Gaposchkin}{1957}]{Payne57} Gaposchkin C. H. P., 1957, The Galactic Novae. (Amsterdam, North-Holland Pub. Co.)

\bibitem[\protect\citeauthoryear{Heywood et al.}{2005}]{Heywood05} Heywood I., O'Brien T.J., Eyres S.P.S., Bode M.F., Davis R.J., 2005, MNRAS, 362, 469

\bibitem[\protect\citeauthoryear{Hjellming}{1979}]{Hjellming79} Hjellming, R. M., 1979, ApJ, 84, 1619

\bibitem[\protect\citeauthoryear{Hjellming}{1996}]{Hjellming96} Hjellming, R. M., 1996, in ASPCS, Vol. 93, Radio Emission from the Stars and the Sun, ed. A. R. Taylor \& J. M. Paredes, 174

\bibitem[\protect\citeauthoryear{H{\"o}gbom}{1974}]{Hogbom74} H{\"o}gbom J. A., 1974, AAPS, 15, 417

\bibitem[\protect\citeauthoryear{Kasliwal et al.}{2011}]{Kasliwal11} Kasliwal M.M., Cenko S.B., Kulkarni S.R., Ofek E.O., Quimby R., Rau A., 2011, ApJ, 735, 94

\bibitem[\protect\citeauthoryear{Kato}{2015}]{Kato15} Kato T. N., 2015, ApJ, 802, 115

\bibitem[\protect\citeauthoryear{Krauss et al.}{2011}]{Krauss11} Krauss M. I. et al., 2011, ApJ, 739, L6

\bibitem[\protect\citeauthoryear{Lang}{1980}]{Lang80} Lang K. R., 1980, Astrophysical Formulae. A Compendium for the Physicist and Astrophysicist. (Springer-Verlag Berlin Heidelberg New York)

\bibitem[\protect\citeauthoryear{Leitherer \& Robert}{1991}]{Leitherer91} Leitherer C., Robert C., 1991, ApJ, 377, 629

\bibitem[\protect\citeauthoryear{Metzger et al.}{2014}]{Metzger14} Metzger B.D., Hasco{\"e}t R., Vurm I., Beloborodov,A.M., Chomiuk L., Sokoloski J.L., Nelson T., 2014, MNRAS, 442, 713

\bibitem[\protect\citeauthoryear{Metzger et al.}{2015a}]{Metzger15} Metzger B. D., Finzell T., Vurm I., Hasco{\"e}t R., Beloborodov A. M., Chomiuk L., 2015, MRNAS, 450, 2739

\bibitem[\protect\citeauthoryear{Metzger et al.}{2015b}]{Metzger15b} Metzger, B.~D., Caprioli, D., Vurm, I., et al.\ 2015, arXiv:1510.07639 


\bibitem[\protect\citeauthoryear{Morlino \& Caprioli}{2012}]{Morlino12} Morlino, G., \& Caprioli, D.\ 2012, \aap, 538, A81

\bibitem[\protect\citeauthoryear{Mukai, Orio \& della Valle}{Mukai et al.}{2008}]{Mukai08} Mukai K., Orio M., della Valle M., 2008, ApJ, 677, 1248

\bibitem[\protect\citeauthoryear{Nagashima et al.}{2013}]{Nagashima13} Nagashima M. et al., 2013, in IAU Symposium, Vol. 281, IAU Symposium, ed. R. Di Stefano, M. Orio, \& M. Moe, 121-123

\bibitem[\protect\citeauthoryear{Nelson et al.}{2014}]{Nelson14} Nelson T. et al., 2014, ApJ, 785, 78

\bibitem[\protect\citeauthoryear{Park, Caprioli \& Spitkovsky}{2015}]{Park15} Park J., Caprioli D., Sptikovsky A., 2015, Phys. Rev. Lett., 114, 085003

\bibitem[\protect\citeauthoryear{Ressler et al.}{2014}]{Ressler14} Ressler, S.~M., Katsuda, S., Reynolds, S.~P., et al.\ 2014, \apj, 790, 85 

\bibitem[\protect\citeauthoryear{Ribeiro et al.}{2014}]{Ribeiro14} Ribeiro V. et al., 2014, ApJ, 792, 57

\bibitem[\protect\citeauthoryear{Roy et al.}{2012}]{Roy12} Roy N., Kantharia N.G., Eyres S. P. S., Anupama G. C., Bode M. F., Prabhu T. P., O'Brien T. J., 2012, MNRAS, 427, L55 

\bibitem[\protect\citeauthoryear{Rybicki \& Lightman}{1979}]{Rybicki79} Rybicki G. B., Lightman A. P., 1979, Radiative Processes in Astrophysics. Wiley-Interscience, New York

\bibitem[\protect\citeauthoryear{Seaquist \& Palimaka}{1977}]{Seaquist77} Seaquist E. R., Palimaka, J., 1977, ApJ, 217, 781

\bibitem[\protect\citeauthoryear{Seaquist et al.}{1980}]{Seaquist80} Seaquist E. R., Duric N., Israel F. P., Spoelstra T. A. T., Ulich B. L., Gregory P. C.,1980, AJ, 85, 283

\bibitem[\protect\citeauthoryear{Shafter}{1997}]{Shafter97} Shafter A. W., 1997, ApJ, 487, 226

\bibitem[\protect\citeauthoryear{Shara et al.}{1997}]{Shara97} Shara M. M., Zurek D. R., Williams R. E., Prialnik D., Gilmozzi, R., Moffat A. F. J., 1997, AJ, 114, 258

\bibitem[\protect\citeauthoryear{Slavin, O'Brien \& Dunlop}{Slavin et al.}{1995}]{Slavin95} Slavin A. J., O'Brien T. J., Dunlop J. S., 1995, MNRAS, 276, 353

\bibitem[\protect\citeauthoryear{Sokoloski, Rupen \& Mioduszewski}{Sokoloski et al.}{2008}]{Sokoloski08} Sokoloski J. L., Rupen M. P., Mioduszewski A. J., 2008, ApJ, 685, L137

\bibitem[\protect\citeauthoryear{Sokoloski et al.}{Sokoloski et al.}{2013}]{Sokoloski13} Sokoloski J.L., Crotts A.P.S., Lawrence S., Uthas H., 2013, ApJ, 770, L33


\bibitem[\protect\citeauthoryear{Staveley-Smith et al.}{Staveley-Smith et al.}{1993}]{Staveley93} Staveley-Smith L., Briggs D. S., Rowe A. C. H., Manchester R. N., Reynolds J. E., Tzioumis A. K., Kesteven M. J., 1993, Nature, 366, 136


\bibitem[\protect\citeauthoryear{Taylor et al.}{1987}]{Taylor87} Taylor A. R., Pottasch S. R., Seaquist E. R., Hollis J. M., 1987, A\&A, 183, 38

\bibitem[\protect\citeauthoryear{Taylor et al.}{1988}]{Taylor88} Taylor A.R., Hjellming R.M., Seaquist E.R., Gehrz R.D., 1988, Nature, 335, 235

\bibitem[\protect\citeauthoryear{Taylor et al.}{1989}]{Taylor89} Taylor A. R., Davis R.J.,  Porcas R.W., Bode M.F., 1989, MNRAS, 237, 81

\bibitem[\protect\citeauthoryear{Williams et al.}{2014}]{Williams14} Williams, B.~J., Borkowski, K.~J., Reynolds, S.~P., et al.\ 2014, \apj, 790, 139 

\bibitem[\protect\citeauthoryear{Wright \& Barlow}{1975}]{Wright75} Wright A. E., Barlow M. J., 1975, MNRAS, 170, 41 

\bibitem[\protect\citeauthoryear{Yamanaka, Itoh \& Komatsu}{Yamanaka et al.}{2010}]{Yamanaka10} Yamanaka M., Itoh R., Komatsu T., 2010, IAU Circ, 9167, 3

\end{thebibliography}
\end{document}